\documentclass[fleqn,usenatbib]{mnras}

\usepackage{newtxtext,newtxmath}
\usepackage[T1]{fontenc}
\usepackage{ae,aecompl}

\usepackage{graphicx}	% Including figure files
\usepackage{xcolor}     % Allows to write coloured text
\usepackage{url}        % Allows to put url in references
\usepackage{hyperref}
\usepackage{comment}
\usepackage{natbib}

\usepackage{caption}
\usepackage{subcaption} % Allows to use subfigures
\title[GWs as inhomogeneity probes]{Mapping the inhomogeneous Universe with Standard Sirens: Degeneracy between inhomogeneity and modified gravity theories}

\author[Kalomenopoulos, Khochfar,  Gair \& Arai]{
Marios Kalomenopoulos$^{1}$\thanks{E-mail: mariok@roe.ac.uk} \href{https://orcid.org/0000-0001-6677-949X}{\includegraphics[scale=0.6]{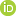}},
Sadegh Khochfar$^{1}$,
Jonathan Gair$^{2}$,
Shun Arai$^{3}$.
\\
$^{1}$Institute for Astronomy, University of Edinburgh, Royal Observatory, Edinburgh EH9 3HJ, UK\\
$^{2}$Max Planck Institute for Gravitational Physics, Potsdam Science Park, Am Muhlenberg 1, D-14476 Potsdam, Germany\\
$^{3}$Center for Gravitational Physics, Yukawa Institute for Theoretical Physics, Kyoto University, Kyoto 606-8502, Japan}

\date{Accepted XXX. Received YYY; in original form ZZZ}

\pubyear{2020}

\begin{document}
\label{firstpage}
\pagerange{\pageref{firstpage}--\pageref{lastpage}}
\maketitle

% Abstract of the paper
\begin{abstract}
	The detection of gravitational waves (GWs) and an accompanying electromagnetic (E/M) counterpart have been suggested as a future probe for cosmology and theories of gravity. In this paper, we present calculations of the luminosity distance of sources taking into account inhomogeneities in the matter distribution that are predicted in numerical simulations of structure formation. In addition, we show that inhomogeneities resulting from clustering of matter can mimic certain classes of modified gravity theories, or other effects that dampen GW amplitudes, and deviations larger than $\delta \nu \sim \mathcal{O}(0.1)\ (99\%\ \rm{C.L.})$ to the extra friction term $\nu$, from zero, would be necessary to distinguish them. For these, we assume mock GWs sources, with known redshift, based on binary population synthesis models, between redshifts $z=0$ and $z=5$. We show that  future GW detectors, like Einstein Telescope or Cosmic Explorer, will be needed for strong constraints on the inhomogeneity parameters and breaking the degeneracy between modified gravity effects and matter anisotropies by measuring $\nu$ at $5 \%$ and $1 \%$ level with $100$ and $350$  events respectively.
\end{abstract}

% Select between one and six entries from the list of approved keywords.
% Don't make up new ones.
\begin{keywords}
	methods: theory -- cosmology: gravitational waves - inhomogeneous universe - modified gravity
\end{keywords}

%%%%%%%%%%%%%%%%%%%%%%%%%%%%%%%%%%%%%%%%%%%%%%%%%%

%%%%%%%%%%%%%%%%% BODY OF PAPER %%%%%%%%%%%%%%%%%%

\section{Introduction}

The standard model of cosmology Lambda-Cold-Dark-Matter ($\Lambda$CDM) is based on the assumption that the universe is homogeneous and isotropic, which is supported on large-scales by precise CMB measurements \citep{2018planck}.  These symmetry hypotheses lead to the well-known  Friedmann-Lema{\^i}tre-Robertson-Walker (FLRW) metric to describe the Universe's geometry. Although the standard model has passed successfully many tests, various ``tensions'' exist \citep{Verde_et_al_2019}, so independent confirmations of its basic assumptions are important.

Tests checking the homogeneity and isotropy of the Universe, based on electromagnetic (E/M) observations, have been proposed and performed in the literature (for example \cite{Clifton_et_al_2008, Busti_et_al_2012, Dhawan_et_al_2018}), showing consistency of the standard model, but require more data (so far simple inhomogeneous models are consistent with homogeneity at $< 2 \sigma$ level).  At the same time, the robustness of some basic $\Lambda$CDM assumptions, like the isotropy  or the spatial curvature of the Universe, have recently been under debate (see \cite{Nielsen_et_al_2016, Rubin_et_al_2016, Colin_et_al_2019, Rubin_et_al_2019, Handley_Will_2019, Valentino_et_al_2020, Efstathiou_et_al_2020, Heinesen_et_al_2020, Migkas_et_al_2020}). Most of the above analyses require good, model-independent distance measurements, which in general, are difficult to obtain.

The first direct detection of GWs with an E/M counterpart (``standard siren''\footnote{Historically, ``Standard Sirens'' refer to every compact binary - BBH, BNS, BH-NS - with or without an E/M counterpart \citep{Schutz_1986, Holz_et_al_2005}. Here we are going to use the term as equivalent to any GW detection that has also independent redshift information.}) \citep{GWCounterpart1} presents an alternative way to study fundamental physics and provide an independent probe for assessing our basic assumptions for the Universe. The simultaneous determination of the luminosity distance, based solely on general relativity (GR), and the localisation of the source from the E/M observation, which was demonstrated for example with the first observation of a binary neutron star \citep{FirstCounterpart}, has opened new possibilities to test our cosmological theories, in a model-independent way, by exploiting the observed distances.

There are already various  proposals of how the luminosity distances inferred from GWs standard sirens observations can be used for cosmology. Examples include the determination of the Hubble parameter \citep{Schutz_1986, Abbott_et_al_2017_Standard_Siren_Ho}, constraints on the cosmological parameters \citep{Cutler_Holz_2009}, the distance duality relation \citep{Hogg_et_al_2020} and measurement of peculiar velocities in the local universe to probe gravity and structure growth \citep{Palmese_Kim_2020}.  Moreover, \citep{Seto_et_al_2001} and \citep{Yagi_et_al_2012} suggest a direct measurement of cosmic acceleration with GW sirens by observing a phase shift in the signal (a deviation from the value that is expected in an FLRW spacetime, can be a sign for deviations from the Cosmological Principle's homogeneity). Further applications of standard sirens include constraints  on cosmic anisotropies, e.g. the dipole anisotropy \citep{Cai_et_al_2018,Lin_et_al_2018}, and the spatial curvature, e.g. \citep{Wei_2018}. 

At the same time, various classes of modified theories of gravity predict that distances inferred from GWs deviate from those of their E/M counterpart. These theories include a number of new degrees of freedom to dynamically describe the late-time acceleration of the Universe \citep[e.g.][]{DEmog, FerreiraCosmoTests}. The additions to GR can introduce several new effects on the propagation of GWs, like a different propagation speed or an amplitude decay, compared to  photons \citep{AnisotropicSaltas, Saltas, Nishizawa, Planck_mass, DEGWs}. However, so far tests with current GWs observations \citep{LigoExtraDGW150914, LigoExtraDGW170817, LigoExtraDGWall}, that investigate a massive graviton and phenomenological models of extra-dimensions, show consistency with GR.

As laid out above, several different physical processes can lead to a reduction in the amplitude of GWs as they propagate through the Universe. In this paper, we investigate the measurement of luminosity distances of GWs propagating through inhomogeneities and under the effects of modified gravity theories. To model the effects of inhomogeneities, different approaches have been taken in the literature \citep[see e.g.][for an overview]{Fleury_2015, Helbig_2019}. Perturbations around FLRW \citep{Bertacca_et_al_2018} show negligible effects for the near future detectors. Here we will concentrate on modelling inhomogeneities with the Dyer-Roeder (DR) relation \citep{DyerRoederDistance} and a modification of it, the modified DR relation (mDR) \citep{MisinterprentingSN}, which approximate better small-scale inhomogeneities. In a previous study \citep{CMYoo_etal_2007ApJ} investigated the impact of inhomogeneities on GWs, concentrating  only on an extreme inhomogeneous (empty) DR universe, with a uniform distribution of same mass objects. The focus of that study was, however, on the information that could be obtained about the lenses, using strong lensing of GW events. Moreover, previous studies \citep{Cutler_Holz_2009, Camera_Nishizawa_2013, Congedo_Taylor_2019} investigated the impact of inhomogeneities in a statistical sense, deriving constraints on the cosmological parameters from joint weak lensing and gravitational waves observations, showing a potential for $\mathcal{O}(1\%)$ accuracy. At the same time, an improved modelling of the lensing magnification distribution, assuming a non-Gaussian PDF \citep{Hirata_et_al_2010, Shang_Haiman_2011} or the extreme case of empty beams \citep{Linder_2008}, can lead to further enhancement in accuracy. The possibility for cosmological constraints from the anisotropy of GWs observations only (without a counterpart) was also studied in \citep{Namikawa_et_al_2016}. In this paper we investigate the possible constraints from future GWs observations, on arbitrary inhomogeneous models and on inhomogeneous models derived from cosmological N-body simulations. We also examine possible degeneracies of these effects with modified gravity theories and how these could be broken with next generation GW detectors, more specifically Einstein Telescope (ET) and Cosmic Explorer (CE)  \citep{Einstein_Telescope, Abbott_2017_next_gen}.

The paper is structured as follows: In Section~\ref{sec:dist} we study the effects on the GW luminosity distance produced by modified gravity theories and an inhomogeneous background and present the various models we are interested in constraining, in Section~\ref{sec:sims} we describe the cosmological simulations and the numerical techniques employed, in Section~\ref{sec:results} we discuss our results and in Section~\ref{sec:conclusions} we summarise our findings and discuss possible future directions. 

\section{Distances from GWs}\label{sec:dist}

A GW detection with an E/M counterpart leads to two observable quantities for cosmology: the luminosity distance to the source\footnote{There is a degeneracy with orbit inclination which can be lifted with precise observations of the two polarisations or E/M observations \citep{Baker_et_al_2019}.} $d^{gw}_L$ and its redshift $z$. The dominant, quadrupole contribution \citep{maggiore1} gives:
\begin{equation}
h = \frac{4}{d^{gw}_L} \left(\frac{G \mathcal{M}_z}{c^2}\right)^{5/3} \left(\frac{\pi f_{\rm gw}}{c}\right)^{2/3},
\end{equation}
where $h$ is the amplitude of the GW, $c$ is the speed of light, $G$ Newton's constant, $f_{\rm gw}(t)$ the frequency of the GW at the observer and $\mathcal{M}_z = (1+z)\mathcal{M}_c$ is the ``redshifted chirp mass'' and $\mathcal{M}_c = (m_1m_2)^{3/5}/(m_1+m_2)^{1/5}$ is the chirp mass, with $m_1$ and $m_2$ the individual masses of the compact objects. For binary black hole systems, the information of redshift is completely degenerate with the chirp mass, and thus a cosmological model, which makes a correspondence between $z$ and $d^{gw}_L$, cannot be constrained by a GW observation alone\footnote{For binary neutron stars, the tidal effects can permit a direct redshift measurement.}. Redshift is inferred by an E/M counterpart, or by a galaxy that is identified as the host of a GW event \citep{Nishizawa:2016ood}. Hereafter, we consider GW events where the redshift $z$ is known. Hence, we consider the case where we have two independent observables, the redshift to the source obtained by E/M observations and its luminosity distance, inferred by the GW detection.

In general, the redshift and luminosity distance have been measured by a number of E/M observations. The measured luminosity distance, $d^{\rm E/M}_L(z)$ is consistent with that of a flat $\Lambda$CDM universe (ignoring radiation):
\begin{equation}\label{eq:LCDM_dist}
d_L^{\rm E/M} (z) = (1+z)\frac{c}{H_0} \int_{(1+z)^{-1}}^1 \frac{dx}{\sqrt{\Omega_m x + \Omega_{\Lambda}x^4}},
\end{equation}
where $\Omega_m,\ \Omega_{\Lambda}$ are the dimensionless densities for matter and the cosmological constant respectively and $H_0$ the Hubble parameter. This fact allows us to assume that the luminosity distance is unique with the underlying cosmological model, i.e., $d^{gw}_L(z) = d^{\rm E/M}$(z). However, this assumption should be carefully inspected with GW observations.

Any deviations between the luminosity distance derived from the GW and the one calculated from the standard $\Lambda$CDM formula, given the redshift of the E/M counterpart, are potential signs of differences caused by systematic deviations from the underlying model assumptions. In the following sections we review the effects caused by modified gravity theories and inhomogeneities in the Universe, and the possibility to constrain them.

\subsection{Modified Gravity}\label{sec:MG}

We start by reviewing the effects of modified gravity that can lead to a decrease on the observed amplitude of a GW\footnote{We note that modified gravity models can also impact the background expansion and hence affect photon propagation \citep{ModifiedStandardSirens, DEGWs}, but we are going to neglect this here.}. We are interested in these, due to their possible degeneracy with the presence of inhomogeneities, that also result in a drop of the observed amplitude (Section \ref{sec:homog_probe}). The new additions to GR can introduce several new effects on the propagation of GWs \citep{AnisotropicSaltas, Saltas, Nishizawa, Planck_mass, DEGWs}. These are model dependent, but effectively they are summarised in the equation below (primes denote derivatives w.r.t. conformal time):
\begin{equation}
h''_{ij} + [2+\nu(\tau)] \mathcal{H}h'_{ij} + [c_T(\tau)^2k^2+a(\tau)^2\mu^2]h_{ij} = a(\tau)^2\Gamma \gamma_{ij},
\end{equation}
where the standard evolution of tensor perturbations is modified by $\nu$ an additional friction term, $c_T$ the GW propagation speed, $\mu$ the graviton mass and $\Gamma \gamma_{ij}$ a source term. The case where $\nu,\ \mu,\ \Gamma$ are equal to zero and $c_T$ = 1 corresponds to standard GR. Focusing only on the terms affecting the amplitude, and neglecting any source terms, we have:
\begin{equation}
h''_{ij} + [2+\nu(\tau)] \mathcal{H}h'_{ij} + k^2h_{ij} = 0.
\end{equation}

The general class of theories that predict a friction term $\nu$ is the class of scalar-tensor theories \citep{Saltas} and theories that break some fundamental assumptions, introducing for example extra-dimensions \citep{Corman_et_al_2020} or theories where non-local terms appear at the level of the quantum effective action \citep{Belgacem_et_al_2018}. Here, because of the friction term in front of the Hubble drag, the distances measured from GWs and E/M signals can differ. Usually the GWs distance is compared to the standard $\Lambda$CDM one - see eq. (\ref{eq:LCDM_dist}) - which is calculated using the redshift obtained from the E/M observation. Then, the relation between the GW and E/M luminosity distance is given by \citep{Belgacem_et_al_2018, Belgacem_et_al_2020}:
\begin{equation}\label{eq:mod_GW_dist_1}
d_L^{\rm gw}(z) = d_L^{\rm E/M}(z) \exp \left\{\int_0^z \ \frac{\nu(z')}{2}\frac{dz'}{1+z'}  \right\} .
\end{equation}
For constant $\nu$, it is easy to solve this relation analytically:
\begin{equation}\label{eq:mod_GW_dist_2}
d_L^{\rm gw}(z) = d_L^{\rm E/M}(z) \ (1+z)^{\nu/2} .
\end{equation}

This case was investigated in \citep{NishizawaFuture}, where the authors found that future constraints on $\nu$ can reach $1 \%$. Similarly, strong observational bounds on scalar-tensor theories have been predicted in \citep{Agostino_Nunes_2019}. The GW170817 event provided a very loose constraint on $\nu$, in the range $-75.3 \leq \nu \leq 78.4$, when considered a constant. The constraints are weaker when $\nu$ is Taylor-expanded as $\nu = \nu_0 - \nu_1 H_0 t_{LB}$, with $t_{LB}$ the look-back time, and $\nu_0,\ \nu_1$ are fitted \citep{AraiHorndeski}. Throughout this paper, we assume that $\nu$ is constant.

\subsection{Inhomogeneous models}\label{sec:homog_probe}

Inhomogeneous models affect both GWs and E/M distances on the same way. This is because they assume that both signals travel in paths that deviate from the smooth background, although without altering GR. A critical investigation of these effects is necessary, since comparing the GWs distance with the one based on $\Lambda$CDM can lead to the false assumption of a deviation from the standard model that can be attributed to a modified GW propagation, while the source of the discrepancy would be a simplified approximation of the E/M distance, when inhomogeneities should be taken into account.

Below we review how different models can affect the angular diameter distance, and hence the luminosity distance that we would infer.

\subsubsection{The FLRW distance}

As a reference, we recall that for a smooth, homogeneous universe, the angular diameter distance $D$, is given from the differential equation \citep{MisinterprentingSN}:
\begin{equation}
\frac{d^2 D}{dz^2} + \left(\frac{d\ln H}{dz}+\frac{2}{1+z}\right)\frac{dD}{dz} =  -\frac{3}{2}\Omega_m\frac{H_0^2}{H^2}(1+z)D,
\end{equation}
where $H(z)^2 = H_0^2[\Omega_m (1+z)^3+\Omega_{\Lambda}+\Omega_k(1+z)^2]$.

\subsubsection{The DR distance}

The Dyer-Roeder (DR) distance models a bundle of rays travelling through voids (``empty beams''), while assuming that the general background follows the standard FLRW geometry. Its usefulness is due to the fact that it yields the largest possible distance (for a given redshift) for light bundles
which have not passed through a caustic \citep{gravitational_lenses}. 

On the other hand, if the universe is not very clumpy, the Dyer-Roeder  distances and those of the smooth FLRW universe are not too different \citep{Fukugita, gravitational_lenses}.

The DR distance is given by the solution to the following differential equation:
\begin{equation}
\frac{d^2 D}{dz^2} + \left(\frac{d\ln H}{dz}+\frac{2}{1+z}\right)\frac{dD}{dz} =  -\frac{3}{2}\Omega_m\frac{H_0^2}{H^2}(1+z)\alpha(z)D,
\end{equation}

where $H(z)^2 = H_0^2[\Omega_m (1+z)^3+\Omega_{\Lambda}+\Omega_k(1+z)^2]$ and only an extra factor $\alpha (z)$ - the inhomogeneity factor - has been added to the RHS compared to the FLRW case. For $\alpha=1$, we return to the standard FLRW result, while $\alpha = 0 $ describes an ``empty'' beam, with the intermediate values denoting different levels of under-dense regions. Values of $\alpha >1$ correspond to over-densities.

\begin{figure}
	\includegraphics[width=\columnwidth]{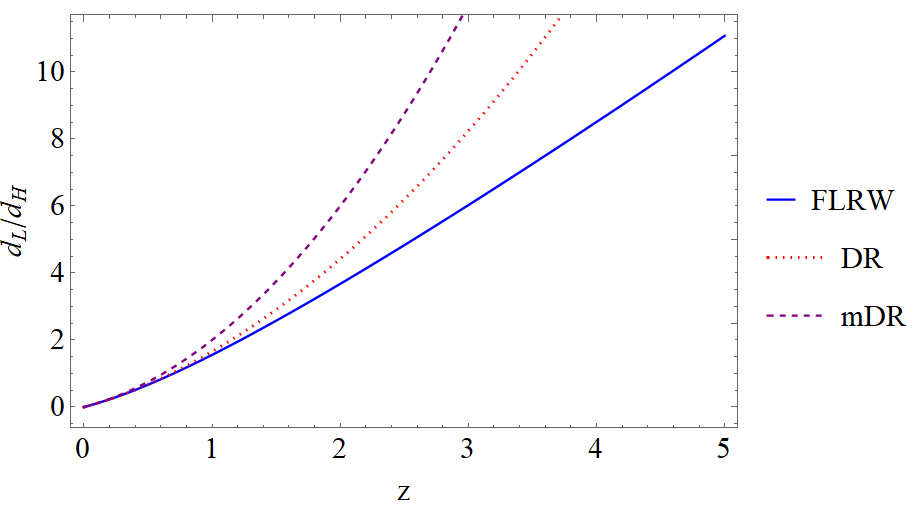}
	\caption{An extreme example (``empty beam'' case, $\alpha=0$) of the differences between the different distant measures. The three models start deviating significantly for $z\geq1$. }
	\label{fig:three_distances}
\end{figure}

\subsubsection{The modified DR distance}

As we mentioned above, the DR approximation assumes that we can disentangle the inhomogeneities encountered by photons from the background density, which assumes a smooth FLRW background. However, photons only experience the local curvature and expansion, so a simple modification would be to take into account the different expansion dynamics caused by the local anisotropies. This leads to a modified version of the DR formula\footnote{Please note, that the original equation in \citep[eq. (82)]{MisinterprentingSN} includes typos and that we show here the correct equation, including $H_0^2$ and a missing factor of two.}:
\begin{align}
\frac{d^2 D}{dz^2} +& \left(\frac{(1+z)H_0^2}{2H^2}[3\alpha(z)\Omega_m (1+z)+2\Omega_k]+\frac{2}{1+z}\right)\frac{dD}{dz} = \nonumber \\ 
& -\frac{3}{2}\Omega_m\frac{H_0^2}{H^2}(1+z)\alpha(z)D,
\end{align}
with $H(z)^2 = H_0^2[\alpha(z)  \Omega_m (1+z)^3+\Omega_{\Lambda}+\Omega_k(1+z)^2]$, where the inhomogeneous parameter $\alpha (z)$, has now been included in all the density terms ($\rho_m \rightarrow \alpha\rho_m$), but its derivatives have been neglected, as in \citep{MisinterprentingSN}.

As can be seen from Figure \ref{fig:three_distances}, the distance of the DR and mDR approximations (in the extreme case of totally empty beams, i.e.  $\alpha=0$) vs the FLRW case can become quite important for high-z observations, and a careful modelling is needed if we want to extract accurate parameters.

Although these redshifts ($z \geq 1$) are large enough to reach homogeneity, in most cases we observe focused light ``beams'', so we usually do not see an average over the whole sky \citep{Weinberg_1976}. Hence, inhomogeneities at relatively large redshifts still need to be accounted for \citep{Fleury_et_al_2017, Dhawan_et_al_2018}.

\subsubsection{The inhomogeneity parameter}

The distances above are not mathematically exact solutions of GR, but  are effective models trying to capture the impact of inhomogeneities on observed distances \citep{gravitational_lenses, Mattson_2010}. Different parameterisations have been suggested in the literature for such effective models. Here we will focus on the following two \citep[][]{Linder_1988, Bolejko_2011}:

\begin{enumerate}\label{eq:parameterisations}
	\item $\alpha(z) = a_0+a_1 z$.
	\item $\alpha(z) = 1 + f(z) \langle \delta \rangle_{1D}$,
\end{enumerate}{}
where $a_0,\ a_1$ are arbitrary constants, the function $f(z)$ is chosen as $f(z) = (1+z)^{-5/4}$ to be consistent with the weak lensing approximation\footnote{This was investigated only for the DR model in \citep{Bolejko_2011}.} \citep{Bonvin_et_al_2006} and $\langle \delta \rangle_{1D}$ denotes the average present-time density contrast along a ray, with $\delta = \delta \rho /\rho$. The two parameterisations are connected at small redshifts via $a_0 = 1 + \langle \delta \rangle_{1D}$ and $a_1 = -5\langle \delta \rangle_{1D}/4$ (see Appendix \ref{sec:ApC} for details).  We obtain the values of $\delta$ from cosmological simulations described in the next section.

\begin{figure}
	\includegraphics[width=\columnwidth]{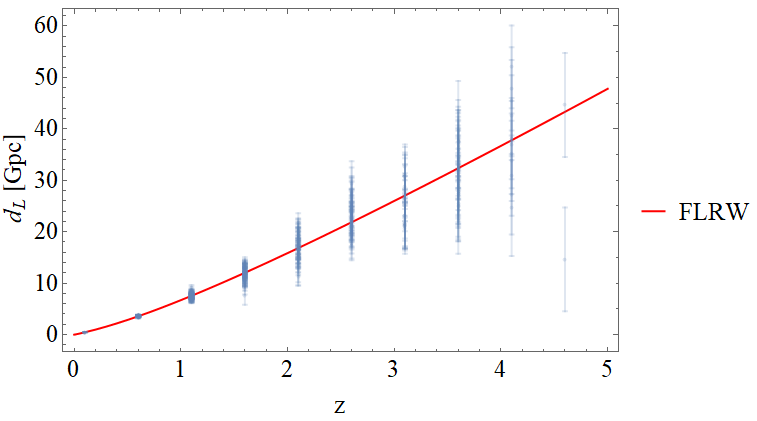}
	\caption{Luminosity distances versus redshift, for a number of mock sources, in redshift bins. An example of how the observational error affects the  ``true'' input distances at a given redshift. In this example we assume an underlying FLRW model.}
	\label{fig:Distances_scatter}
\end{figure}

\subsection{Constraining the inhomogeneity parameter}\label{sec:HomPar}

The differential equations above give the angular diameter distance in each specific case. We solve them, using as initial conditions $D(z=0)=0$ and $ D'(z=0) = c/H_0$, and assuming, for consistency with the later sections, the cosmological parameters of WMAP9 \citep{Hinshaw_et_al_2013_WMAP}, with ($\Omega_m,\ \Omega_{\Lambda},\ \Omega_k,\ H_0) = (0.285,\ 0.715,\ 0,\ 69.5$).

The choice of priors on the standard cosmological parameters, based on the concordance cosmological model, let us proceed to constrain the ``clumpiness'' of the universe alone, or else the fraction of DM in compact objects. This has been attempted before, in a somewhat more simplified way (with constant $\alpha$), for Supernovae (SN) and Gamma-ray Bursts (GRBs) in \citep{Breton_et_al_2013, Helbig_2015a, Helbig_2015b}, however our work extends the analysis to higher redshifts, uses the full non-linear clustering  found in cosmological N-body simulations and introduces different systematics, related to the GW observations\footnote{Previous studies \citep{Basti_et_al_2012, Fleury_et_al_2013, Dhawan_et_al_2018} that have fitted simultaneously the cosmological parameters and $\alpha$ do not find a significant change of the standard values of $\Omega_m$ and $\Omega_{\Lambda}$ - see also Appendix \ref{sec:ApCosmoA}.}.

To transform to luminosity distances we use the reciprocity distance relation \citep{Etherington_1933, Ellis_2007}, which holds for an arbitrary spacetime geometry, as long as photons are conserved, and connects luminosity and angular diameter distances via:
\begin{equation}\label{eq:Etherington}
d_L = (1+z)^2 D
\end{equation}
Concerning the errors in distance determination, we follow the analysis of \citep{Cai_et_al_2017}. We neglect the errors of the spectroscopic redshift determination, since they are negligible compared to the errors in the luminosity distance. For the errors in the GW luminosity distance we take into account the uncertainty from instrumental errors \citep{T_Li_Thesis}:
\begin{equation}
\sigma^{inst}_{d_L} = \frac{2d_L}{\rho(z, M_c)},
\end{equation}{}
where the Signal-to-Noise Ratio (SNR) $\rho(z, M_c)$, depends on the chirp mass of the source, its redshift and the detector's sensitivity, and the factor of $2$ comes from marginalising over the uncertainty of the inclination's determination. Also, there is an additional error, connected to weak lensing effects\footnote{This formula is extrapolated from SN observations \citep{Jonsson_et_al_2010}. Theoretical models, based on numerical simulations \citep{Holz_Linder_2005, Marra_et_al_2013, Fleury_et_al_2015}, each one using different techniques (including inhomogeneous models) are consistent with each other, and the one used above, within observational errors.} \citep{Sathyaprakash_et_al_2010, Zhao_et_al_2011}: 
\begin{equation}
\sigma^{lens}_{d_L} = 0.05zd_L.
\end{equation}{}

Finally, the total error (for a single event) is:
\begin{equation}\label{eq:total_error}
\sigma_{d_L} = \sqrt{\left(\frac{2d_L}{\rho(z, M_c)}\right)^2+(0.05zd_L)^2}
\end{equation}{}

We add these errors in quadrature, assuming they originate from independent sources, as one part is systematic and the other statistical. In the following, when calculating the $\chi^2$ fit, in order to estimate the errors we are going to consider the luminosity distance of the ``true'' case as their source. We assume that all sources are independent, and include a scatter in their distances, allowing them to follow a normal distribution $\mathcal{N}(d_L, \sigma)$, with $d_L$ the mean and $\sigma$ the standard deviation, for each redshift, as shown in Figure \ref{fig:Distances_scatter}. We describe in more detail the calculation of $\rho$ in the next subsection.

\begin{figure}
	\includegraphics[width=\columnwidth]{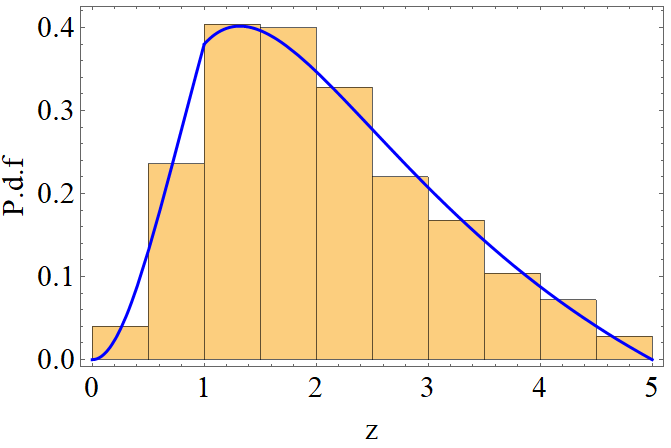}
	\caption{The distribution of our random mock data (histogram), compared to the theoretical distribution. A $\Lambda$CDM cosmology (WMAP9 - see text) is assumed.}
	\label{fig:sources_distribution}
\end{figure}

\subsection{SNR calculation}\label{sec:SNR_calculation}

We estimate the SNR as:
	\begin{equation}
	\rho^2 = 4\int_0^{f_{max}} df \frac{|\tilde{h}(f)|^2}{S_n(f)},
	\end{equation}
	where $\tilde{h}$ is the Fourier transform (FT) of the amplitude of the GW, $S_n(f)$ is the characteristic noise curve of each detector (see Appendix \ref{sec:ApSourceSimulator}) and $f_{max} = 2 f_{ISCO}$ is two times the frequency at the innermost stable circular orbit, where the analytical methods fail.
	
	For the FT of the amplitude of a binary system we follow \citep{buonanno2007gravitational, maggiore1} and calculate the sky-averaged value, over all possible directions and inclinations. Since we are only interested in the amplitude's norm, the relevant phase factors are simplified. Then, for an L-shaped detector, like LIGO or CE, we have:
	\begin{equation}
	|\tilde{h}(f)| = \sqrt{2/15} \pi^{-2/3}G^{5/6}c^{-3/2}\mathcal{M}_z^{5/6}f^{-7/6}d_L(z)^{-1}.
	\end{equation}
	For a V-shaped detector, like ET, with an opening angle of $60$ degrees and three arms \citep{SNR_ET_Regimbau_et_al_2012}, we find only a difference in the normalisation coefficient due to the different detector configuration. This leads to the $\sqrt{2/15}$ factor being substituted by $\sqrt{3/10}$ in above equation.

\subsection{Merger Rates}\label{sec:MR}

The redshift distribution of potentially observable GW sources (Figure \ref{fig:sources_distribution}) is given by \citep{Zhao_et_al_2011}:
\begin{equation}\label{eq:rate}
P(z) \sim \frac{4 \pi D^2_c(z) R(z)}{H(z)(1+z)}, 
\end{equation}
where $D_c(z)$ is the comoving distance and R(z) the merger rate (number of events per year) of compact binary systems with E/M counterparts (BH-NS $\&$ NS-NS) given by \citep{Schneider_et_al_2001, Cutler_et_al_2009, Zhao_et_al_2011}:
\begin{equation}
R(z)=
\begin{cases}
1+2z,\ z \leq 1\\
\frac{3}{4}(5-z),\ 1<z<5\\
0,\ z \geq 5.
\end{cases}
\end{equation}
The rates are based on the binary population synthesis code \verb#SeBa# \citep{Portegies_et_al_1998}, where the main assumptions used for the calculation in \citep{Schneider_et_al_2001} are: (i) the mass
of the primary star $m_1$ is determined using the mass function described by \citep{Scalo_1986} between $0.1$ and $100\ M_{\odot}$, (ii) For a given $m_1$, the mass of the secondary star $m_2$ is randomly selected from a uniform distribution between a minimum of $0.1\ M_{\odot}$ and the mass of the primary star, (iii) The semi-major axis distribution is taken flat in $log(a_{\rm{axis}})$ ranging from Roche-lobe contact up to $10^6\ R_{\odot}$ and (iv) The initial eccentricity distribution is independent of the other orbital parameters. The results are constrained by our Galaxy, but, using the observed star formation rate, they are extended until redshift $z=5$.

In the rate equation - eq. (\ref{eq:rate}) - we can assume, without affecting our results, the estimates for a smooth FLRW cosmology, since we are interested in the deviations of distances between some observed sources. This should not be affected by how we populate the sources (see also Appendix \ref{sec:ApA}). The evolution of merger rate as a function of redshift has also been constrained from the up-to-date detections \citep{LigoPopulationRates}, but with poor results so far due to the small statistics. Only the presence of a significant number of lensed events could change considerably the distribution, however this does not seem to be the case \citep{Hannuksela_et_al_2019}. Nevertheless, the method proposed in this paper is not invalidated by the distribution of sources, however the latter is important for the estimated error contours.

\begin{figure}
	\includegraphics[width=\columnwidth]{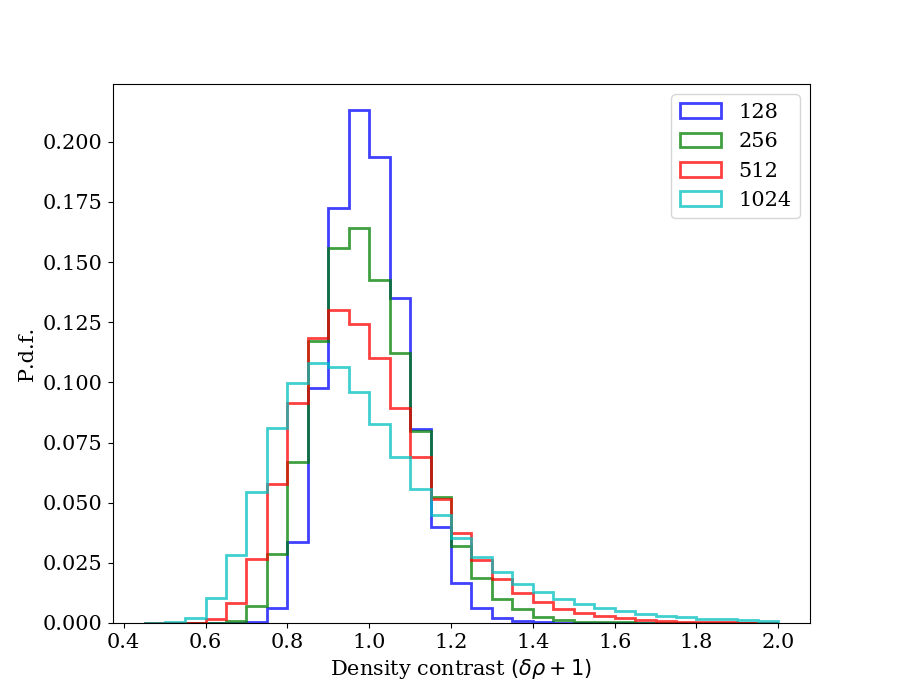}
	\caption{Distribution of densities along 1D trajectories $(\langle \delta \rangle_{1D} + 1)$ on simulations of different resolution, at $z=0$. The higher resolution run, which corresponds to a density averaging in $\sim 8\ \rm{Mpc}^3$ cubes gives the more interesting tails, being able to resolve better the small-scale structure.}
	\label{fig:Density_contrast}
\end{figure}

\section{Simulations}\label{sec:sims}

To calculate realistic density anisotropies, we  rely on cosmological dark matter only simulations run with \emph{Gadget-4} \citep{Gadget_4} for the LEGACY project.

The latter is composed by two primary volumes of $1600$ Mpc/h and $100$ Mpc/h box sizes run down to $z=0$ with $2048^3$ resolution elements, as well as a set of zoom-in simulations on the larger box, with size $83$ Mpc/h, and an effective resolution of $32768^3\ (1700^3)$. These simulations have been designed to sample $-2, -1, 0, +1$, and $+2$ $\sigma$ of the mean density value as well as extremely high (cluster) and very low (void) density regions and are therefore ideal to study different environmental effects.

For our analysis, we use the data from the big $1600$ Mpc/h run, at $z=0$, with an effective resolution of $2048^3$ particles and $5.43 \times 10^{10} M_{\sun}$ mass resolution which will suffice for the purpose of this investigation.

\begin{figure*}
	%\addtocounter{figure}{-1}
	\centering
	\begin{subfigure}[b]{0.42\textwidth}
		\centering
		\includegraphics[width=\textwidth]{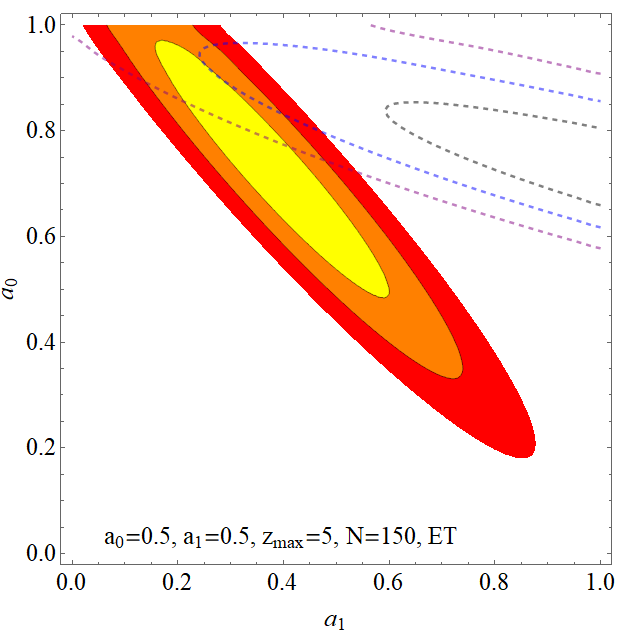}
	\end{subfigure}
	\hspace{1cm}
	\begin{subfigure}[b]{0.42\textwidth}
		\centering
		\includegraphics[width=\textwidth]{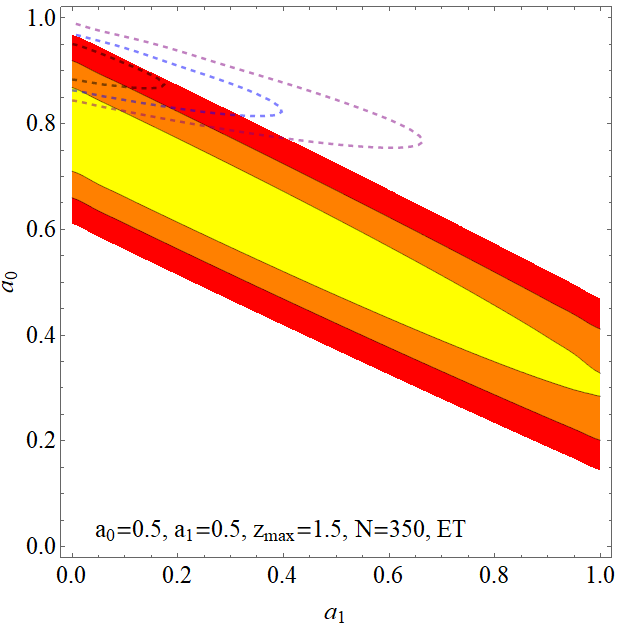}
	\end{subfigure}
	\newline
	\begin{subfigure}[b]{0.42\textwidth}
		\centering
		\includegraphics[width=\textwidth]{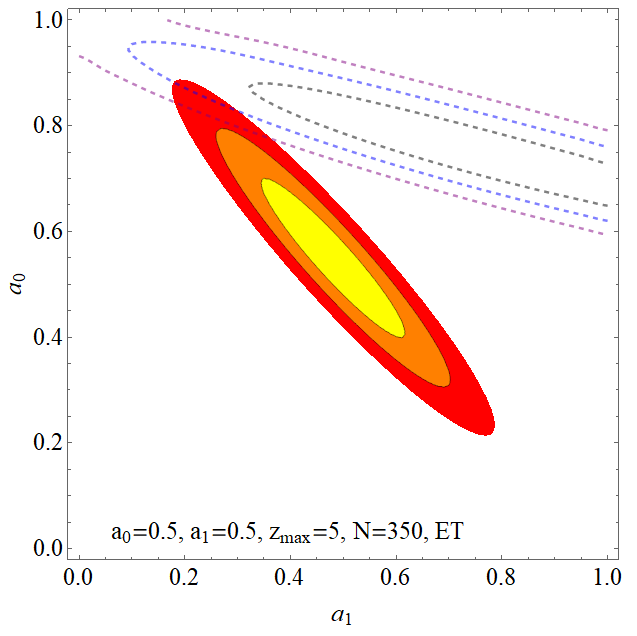}
	\end{subfigure}
	\hspace{1cm}
	\begin{subfigure}[b]{0.42\textwidth}
		\centering
		\includegraphics[width=\textwidth]{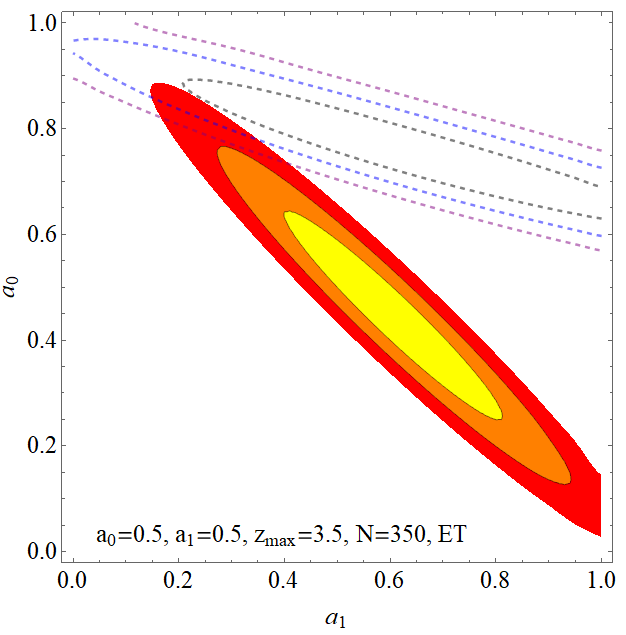}
	\end{subfigure}
	\hspace{1cm}
	\begin{subfigure}[b]{0.42\textwidth}
		\centering
		\includegraphics[width=\textwidth]{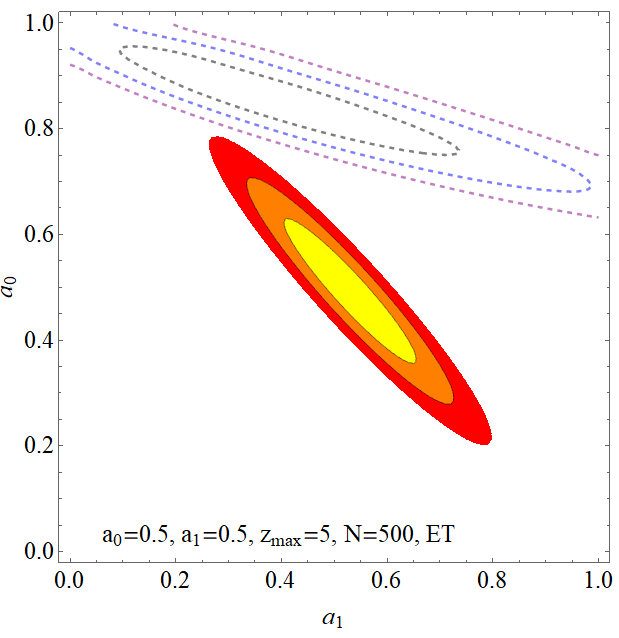}
	\end{subfigure}
	\hspace{1cm}
	\begin{subfigure}[b]{0.42\textwidth}
		\centering
		\includegraphics[width=\textwidth]{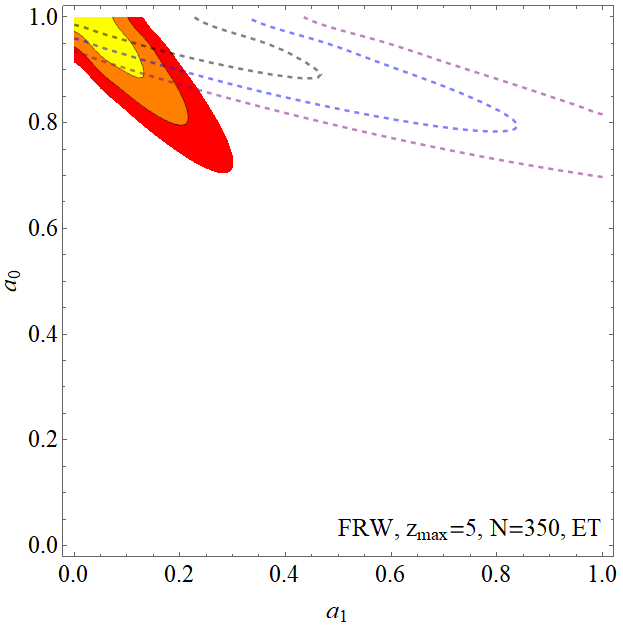}
	\end{subfigure}
	\caption{The left column shows how the constraining power of standard sirens changes with number of sources. The two upper figures in the right column show the effects of maximum redshift, when total number of sources is held fixed.The input distances in these cases follow a DR model with $(a_0,\ a_1) = (0.5,\ 0.5)$. The bottom right plot shows a ``sanity check'' fit to an FLRW input. Filled contours show constraints on the DR models, while dashed lines on the mDR ones.}
	\label{fig:Scaling_plots}
\end{figure*}

\subsection{Cloud-in-cells implementation}

For our simulation box, we perform a Cloud-in-Cell (CiC) interpolation scheme to deposit the particles to specific grid points \citep{hockney1988computer}. The weighting function is:
\begin{equation}
W(x-x_p) = 
\begin{cases}
1 - \frac{|x-x_p|}{L},\quad |x-x_p| \leq L\\
0,\quad \textrm{otherwise},
\end{cases}
\end{equation}
where $L$ is the resolution of the CiC box and $x_p$ the centres of the grid cells.

\subsubsection{Density calculation}

We investigate a number of resolutions, from a grid with $128^3$ cells to a grid with $1024^3$  cells. From these we calculate the density anisotropies along straight trajectories and calculate the mean value for each ray\footnote{See Appendix \ref{sec:ApRays} for details.} to find $\langle \delta \rangle_{1D}$ (Figure \ref{fig:Density_contrast}). For all resolutions the mean 3D density contrast $\delta \rho$ is zero.

As expected, the lowest resolution ($128^3$) returns a distribution closer to the mean contrast, since we average out on larger volumes. The highest resolution run ($1024^3$) produces a distribution with tails that describe the under/over-densities at small scales. In the following we exploit the data from this run.

\section{Results}\label{sec:results}

\subsection{Inhomogeneity Constraints}\label{sec:CosmoPar}

To fit our model with GW standard sirens observations we use a $\chi^2$ goodness of fit defined as:
\begin{equation}
\chi^2 = \sum \frac{[d_L(z)-d_L^{\rm gw}(z)]^2}{\sigma_{d_L}(z)^2},
\end{equation}

where the summation includes all distinct observations. We use here $350$ mock GWs events, unless stated otherwise. This corresponds to a conservative limit for future detectors \citep{Abbott_2017_next_gen, Einstein_Telescope} for redshifts $z \geq 2$ per year cycle, but seems to be a quite optimistic limit for a $5$-year LISA mission \citep{LisaCounterpart} and the next observing run of the current GWs detectors \citep{2018Prospects}. For this reason, we are going to use the sensitivity curve of Einstein Telescope (ET) \citep{Einstein_Telescope} - the most conservative between the future generation GW detectors, as our standard reference when investigating future constraints. Current detectors will not be able to put any useful constraints. We return to this at the end of the next subsection.  $d_L^{\rm gw}$ denotes the distance based on the inhomogeneous or modified gravity models we study. For each ``fitted'' distance, the $\alpha$ parameters take all the values between $0$ to $1$ when constraining an inhomogeneous model and $\nu$ varies between $-1$ to $1$ for the modified gravity parameterisation,  based on a standard grid sampling method, with resolution chosen for convergence, and where for each point a different value for $\chi^2$ is calculated. The redshift of the sources are randomly drawn from the distribution in Figure \ref{fig:sources_distribution}, where the maximum possible redshift is chosen as $z_{max}$. We denote as $d_L(z)$ the ``true'' distance, that would be specified in each case (see below), and with this we calculate the errors using eq. (\ref{eq:total_error}).

More specifically, the steps we take are the following:

\begin{enumerate}
	\item We choose the number of sources we want to use, $N$.
	\item We choose the maximum redshift of the sources, $z_{max}$.
	\item We generate an array of redshifts of length $N$ and maximum value $z_{max}$. If we use the densities calculated from the simulation, we similarly generate an array of $\delta$s, randomly drawn from the high-resolution distribution of Figure \ref{fig:Density_contrast}.
	\item We select our ``true'' distance from either the DR, mDR, FLRW or modified gravity model and we transform the redshifts generated above to a list of distances.
	\item We finally calculate the constraints that can be put on the inhomogeneity/modified gravity parameters, by fitting with a theoretical model ($d_L^{\rm gw}$ above).
\end{enumerate}

In our contour fits we plot the $1\sigma, 2\sigma, 3\sigma$ confidence intervals. The filled ones correspond to the DR model and the dashed ones to the mDR model.

\begin{figure}
	\includegraphics[width=\columnwidth]{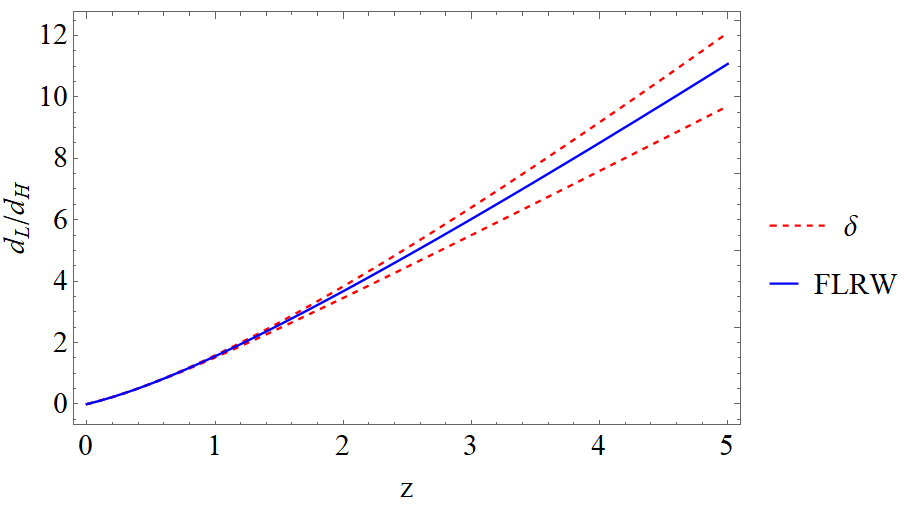}
	\caption{An example of the most extreme expected deviations from FLRW when using the $\delta$ parameterisation, based on simulation data. Values higher (lower) than the FLRW one, correspond to mean under (over) density. A dimensionless distance ratio is plotted, with $d_H$ being the ``Hubble distance'', $c/H_0$.}
	\label{fig:Density_contrast_effects}
\end{figure}

\subsubsection{Scaling with sources and redshift}

We begin by investigating how the constraints scale with number of sources and with maximum value of redshift. For this we simulate $150,\ 350$ and $500$ sources and investigate readshifts $z = 1.5,\ 3.5,\ 5$. The filled contours correspond to contraints based on the DR distance and the dashed ones on the mDR distance, both using the first parameterisation of section \ref{eq:parameterisations} ($\alpha(z) = a_0 + a_1 z$).

Our results are summarised in Figure \ref{fig:Scaling_plots}. We start by testing how well we recover an inhomogenous DR model with input parameters $(a_0,\ a_1) = (0.5,\ 0.5)$ and a FLRW model. First of all, in both cases we recover the input models at the $1\sigma-$level and in the latter case the model is constrained significantly.

Secondly, we observe a similar trend as \citep{MisinterprentingSN}, in that different inhomogeneous models can lead to quite distinct results as shown by the dashed-line contours compared to the solid line ones. These uncertainties in the modelling, limit precision cosmology in terms of being able to constrain physical properties (e.g. the optical depth), modified gravity theories or parameterisations of the dark energy equation of state to high accuracy \citep{Chevallier_et_al_2001, Linder_2003}. 

Thirdly, as expected, we observe that both parameterisations lead to more stringent constraints when more events are detected. Also, high-z observations are required to better distinguish between the models, being consistent with Figure \ref{fig:three_distances}. The two parameters $(a_0,\ a_1)$ are almost unconstrained in the DR case, while $a_0$ can reach $\sim 20 \%$ accuracy if a mDR model is considered. However, the latter seems almost unaffected by either $z_{max}$ or the number of sources.

A similar conclusion results from the analysis of $350$ GW sources, but for different maximum redshifts $z_{max} = 1.5,\ 3$. The relevant figures demonstrate clearly that the maximum redshift is a much more important parameter than the number of sources, since the different distance measures start to deviate significantly at larger redshifts. This is true even if the underlying population synthesis model leads to quite similar distribution of SNR per maximum redshift - see Appendix \ref{sec:ApSourceSimulator}. This would allow a direct probe of the scales where the Universe reaches homogeneity, especially if the errors can be reduced by de-lensing techniques \citep{Lewis_et_al_2006}. 

We should note here that we repeated a similar procedure with current detectors (aLIGO). From the $350$ sources that were observed by ET, only $28$ passed the trigger limit of $\rho=8$. As a result, in all cases current detectors leave the whole parameter space unconstrained. This emphasises the need for new ground-based detectors for cosmological studies.

\subsubsection{Constraints from numerical simulations}

We now follow the same procedure exploiting the second parameterisation in \ref{eq:parameterisations}. This has a clearer physical interpretation, related directly with the density anisotropies along the line-of-sight. Here we keep the maximum redshift and the number of sources fixed at $z_{max} = 5,\ N=350$, and consider as ``fiducial'' distances the ones based on the $\delta$ parameterisation ($\alpha(z) = 1 + f(z) \langle \delta \rangle_{1D}$).

To calculate them, we follow the following procedure: For each mock source, we pick randomly a 1D density contrast from the high resolution distribution of Figure \ref{fig:Density_contrast} and a random redshift as before. We calculate the ``true'' distance by numerically solving the DR equation, since in this model this would be equivalent to the weak lensing approximation. With them, we try to constrain the DR and mDR models, based on the first parameterisation of section \ref{eq:parameterisations} ($\alpha(z) = a_0 + a_1 z$).

\begin{figure}
	\includegraphics[width=\columnwidth]{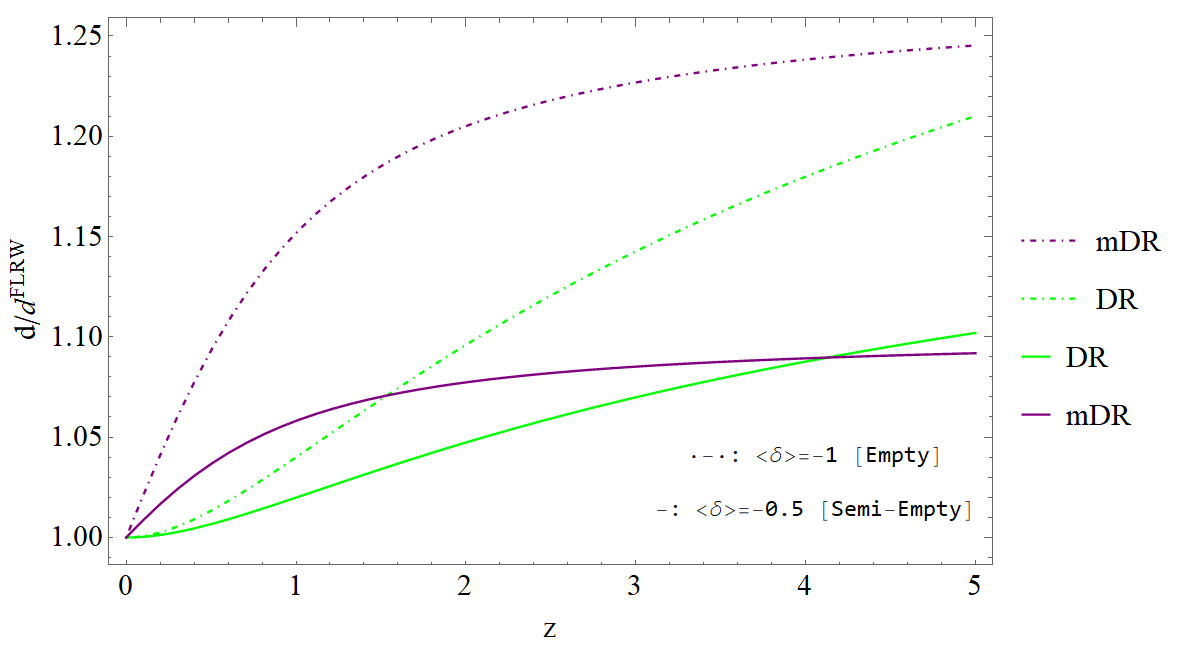}
	\caption{Ratio of modified distances over the FLRW one for two cases of big under-densities: ``empty'' case, $\langle \delta \rangle_{\rm 1d} = -1$ (dot-dashed lines) and ``semi-empty'' case, $\langle \delta \rangle_{\rm 1d} = -0.5$ (solid lines).}
	\label{fig:density_ratios}
\end{figure}

\begin{figure}
	\includegraphics[width=\columnwidth]{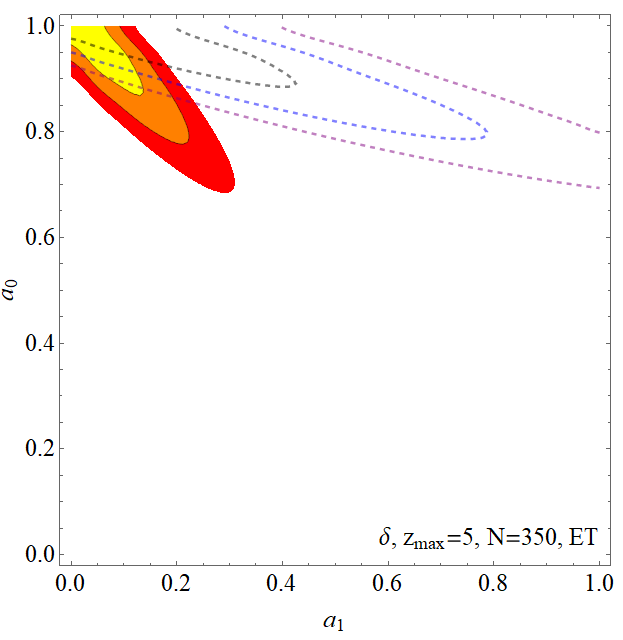}
	\caption{Constraints on the inhomogeneity parameters ($a_0, a_1$) (as above) based on a realistic density distribution from numerical simulations. For both cases there is consistency with a FLRW background.}
	\label{fig:delta_constraints}
\end{figure}

An example of the possible deviation is shown in Figure \ref{fig:Density_contrast_effects}, where we have used the two limiting cases ($\pm 3 \sigma$) based on our simulations. Since these correspond to small perturbations in the metric, the effects are small. More extreme cases are shown in Figure \ref{fig:density_ratios}. The constraints on $(a_0,\ a_1)$ are shown in Figure \ref{fig:delta_constraints}. We see that the presence of small inhomogeneities along the ray results in distance estimates consistent with FLRW, confirming some previous semi-analytical and numerical studies \citep{Mortsell_2002, Kaiser_et_al_2016, Adamek_et_al_2019}. However, we note that the result only holds for weak inhomogeneities\footnote{Note that the maximum mean under-density we find in our simulations is of the order of $\delta \rho/\rho \sim -0.4$.}, since non-linear effects could potentially have an important contribution \citep[e.g.][]{Bolejko_2018}.

\subsection{Modified gravity effects}\label{sec:MoG_const}

\begin{figure*}
	%\addtocounter{figure}{-1}
	\centering
	\begin{subfigure}{0.48\textwidth}
		\centering
		\includegraphics[width=\textwidth]{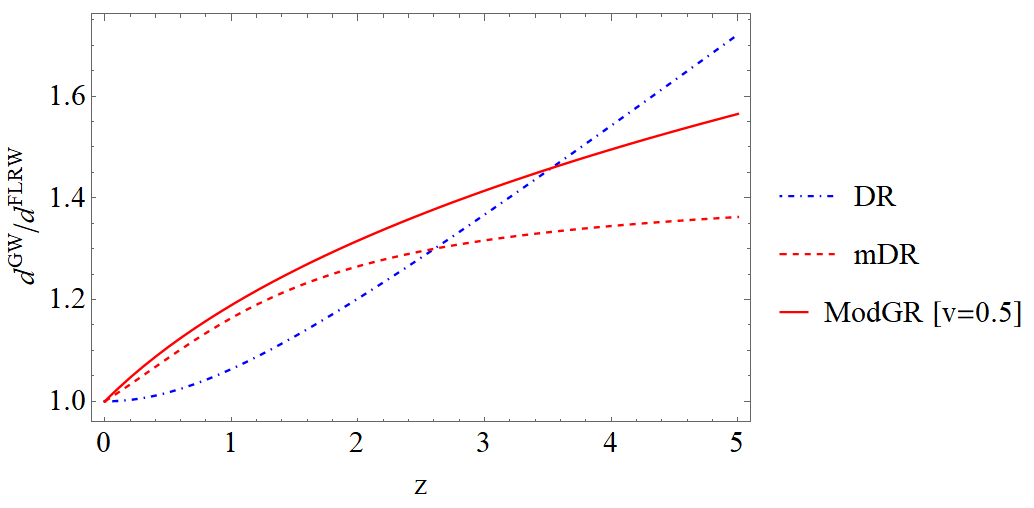}
	\end{subfigure}
	\hfill
	\begin{subfigure}{0.48\textwidth}
		\centering
		\includegraphics[width=\textwidth]{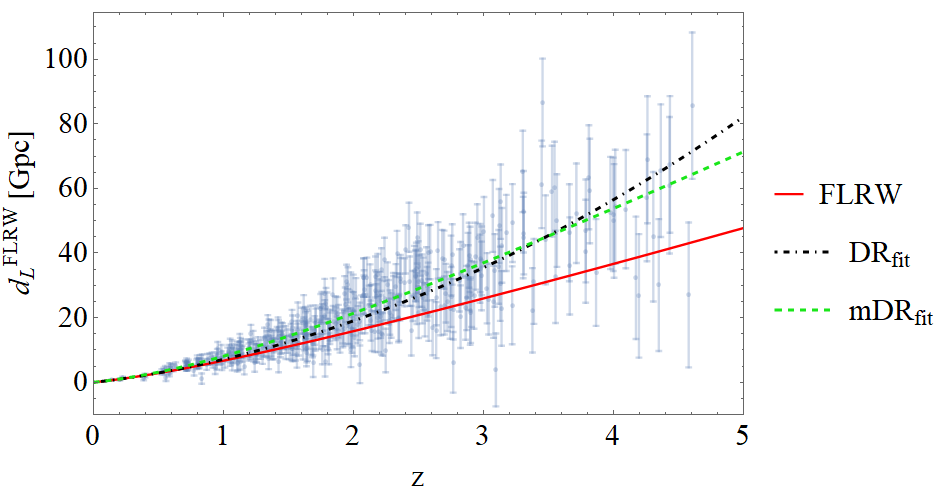}
	\end{subfigure}
	\caption{A comparison of the effect on the GW distances between inhomogeneous models and a general modified gravity parameterisation. Constant, best-fit parameters are $(a_0,\ a_1) = (0,\ 0)$, $(a_0,\ a_1) = (0.2,\ 0.1)$ for DR and mDR models respectively, and the data points were produced to follow a modified gravity with  $\nu=0.5$.}
	\label{fig:MoG_fits}
\end{figure*}

\begin{figure*}
	%\addtocounter{figure}{-1}
	\centering
	\begin{subfigure}[b]{0.433\textwidth}
		\centering
		\includegraphics[width=\textwidth]{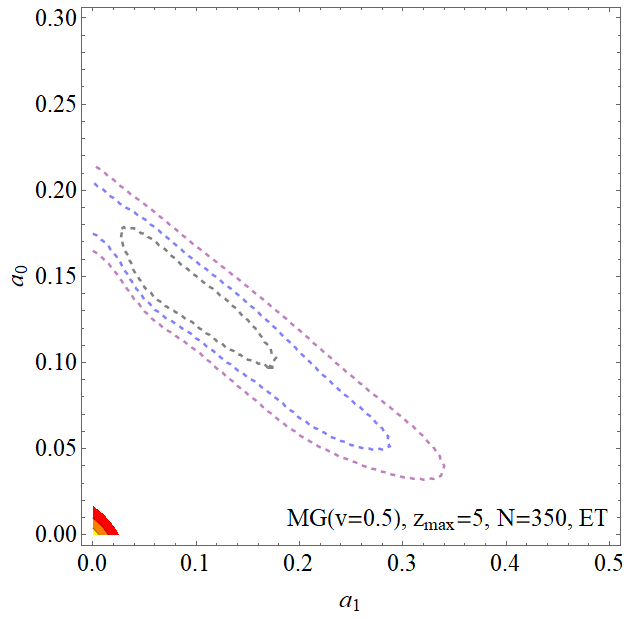}
	\end{subfigure}
	\hspace{1cm}
	\begin{subfigure}[b]{0.43\textwidth}
		\centering
		\includegraphics[width=\textwidth]{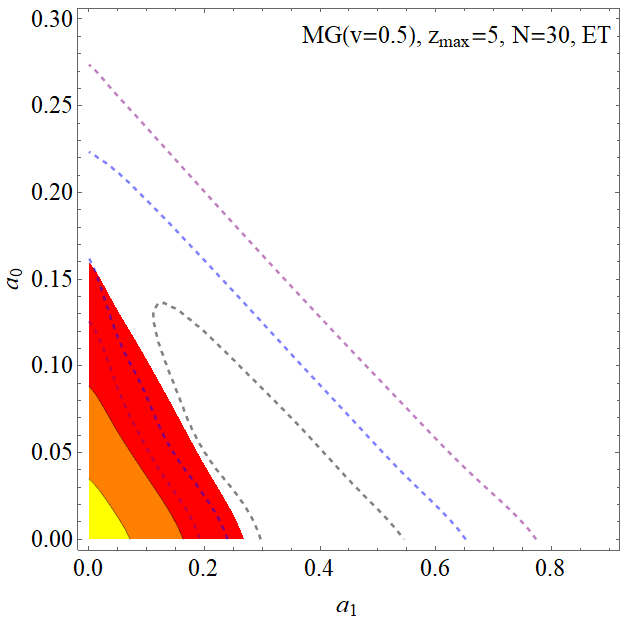}
	\end{subfigure}
	\caption{Contour fits for the two inhomogeneous distance models, when the ``true'' distances follow a modified gravity model with $\nu=0.5$. Next generation detectors, like ET, will be able to break the degeneracies between these models, especially for more extreme cases of their parameters, even with a small number of observations. Note the different ranges in the axes.}
	\label{fig:MoG_constraints}
\end{figure*}

\begin{figure*}
	%\addtocounter{figure}{-1}
	\centering
	\begin{subfigure}{0.49\textwidth}
		\centering
		\includegraphics[width=\textwidth]{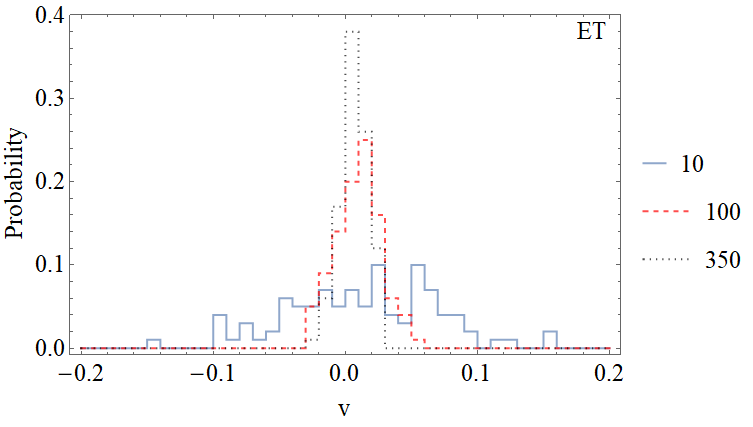}
	\end{subfigure}
	\hfill
	\begin{subfigure}{0.49\textwidth}
		\centering
		\includegraphics[width=\textwidth]{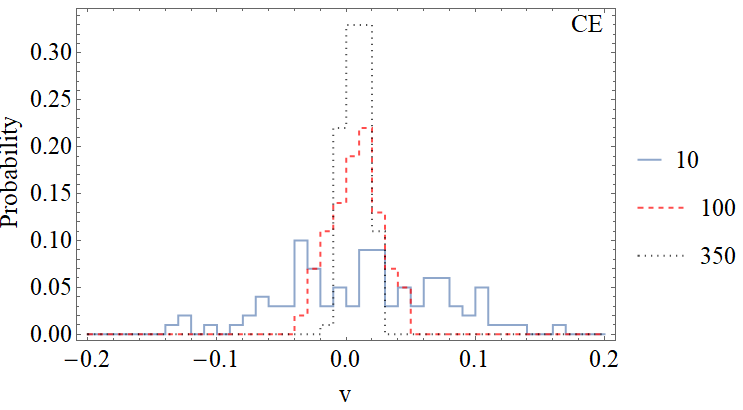}
	\end{subfigure}
	\caption{A low number of observations still leaves space for ``confusion'' between matter anisotropies and modified gravity effects, but as these grow, future detectors are able to reach an accuracy on $\nu$ of the order of $\mathcal{O}(0.01)$ breaking the degeneracy.}
	\label{fig:ET_CE_nu_likelihood}
\end{figure*}

As we have seen, the measurement of accurate distances from GWs can be in itself an independent probe of the inhomogeneity of our Universe. Inhomogeneous models with under-densities, in general predict larger distances than the FLRW ones and this effect could have more general implications, since every physical mechanism that modifies the amplitude of the wave, could lead to a similar deviation \citep{DEGWs, Belgacem_et_al_2018, NonLocalDistance, Wei_2019, Zhou_et_al_2019}. Figure \ref{fig:MoG_fits} provides a simple example of this correspondence.

Of course, an inhomogeneous universe will lead to a different propagation than FLRW, for both photons and GWs, however this strengthens our point that an accurate determination of the underlying geometry is necessary before investigating deviations of modified gravity models, which are usually compared with the expectation from FLRW. This could lead to important implications if the underlying geometry is not FLRW, since in this case modified gravity effects are degenerate with respect to inhomogeneous models.

However, as can  been seen in the example of Figures \ref{fig:MoG_fits}, precise distance measurements, that would significantly reduce the observed errors, could disentangle the two effects in simple models, since they have a different redshift dependence (concave vs convex curves). Also it is worth noticing that inhomogeneous models could possibly \emph{only} mimic gravity modifications that lead to increased distance, so with $\nu > 0$. Hence, although the degeneracy between inhomogeneities and modified gravity models concern mainly small, positive values of $\nu$, that will be investigated in more detail below, since the sign of $\nu$ and even its redshift dependence $\nu (z)$ are not presently constrained significantly, we cannot \emph{a priori} break this degeneracy.

For a more detailed comparison, we repeat our $\chi^2$ fit, where in this case we use as ``true'' input a modified gravity model with effective parameter $\nu=0.5$. We then constrain the values of $(a_0,\ a_1)$ for the two inhomogeneous models, using $350$ mock observations that reach $z_{max}=5$. This leads to Figure \ref{fig:MoG_constraints}. We see that even a small number of sources ($N=30$) results in quite strong constraints, indicating that only extreme inhomogeneities can lead to equivalent results. Of course, this is an arbitrary example, but it demonstrates our previous point that ``realistic'' inhomogeneous models are not able to mimic all values of the modification parameters. The best fit parameters lead to distances as shown in Figure \ref{fig:MoG_fits}, which deviate significantly from the FLRW ones and are probably unphysical. However, modified gravity models that result in less extreme values of $(a_0,\ a_1)$ would not be distinguishable from inhomogeneous models given present observational facilities. 

Finally, we invert the procedure and exploiting the $\delta$ parameterisation we try to fit the best modified gravity model and put limits to the values of $\nu$ that could be disentangled from small inhomogeneity effects. We demonstrate this effect and quantify the number of events needed for better convergence in Figure \ref{fig:ET_CE_nu_likelihood}, where we draw the likelihood of the $\nu$ parameter, for different numbers of events. As can be seen, deviations bigger than $\sim 0.1$ from the FLRW value ($\nu=0$) are needed, in order to be possible to disentangle the modified gravity effect from inhomogeneities, at least for a large number of observations. This shows that the intrinsic scatter of small inhomogeneities is not significant enough  to manage to mimic large deviations from the standard GR case.
	
At the same time, a small number of, low quality, observations can lead to serious misidentifications of inhomogeneity effects with deviations from GR.  At least $100$ standard sirens would be needed for convergence to the ``real'' value with about $\delta \nu \sim \pm 0.05$ accuracy. To reach an order of about $1 \%$ we need at least $350$ sources with counterparts.

\subsection{Next generation GWs detectors}\label{sec:next_gen}

As a final step, we compare some of the results above with another future GW detector, Cosmic Explorer. CE largely overlays the frequencies probed by ET with about half a magnitude better SNR at its more sensitive region.
	
\begin{figure*}
	%\addtocounter{figure}{-1}
	\centering
	\begin{subfigure}{0.48\textwidth}
		\centering
		\includegraphics[width=\textwidth]{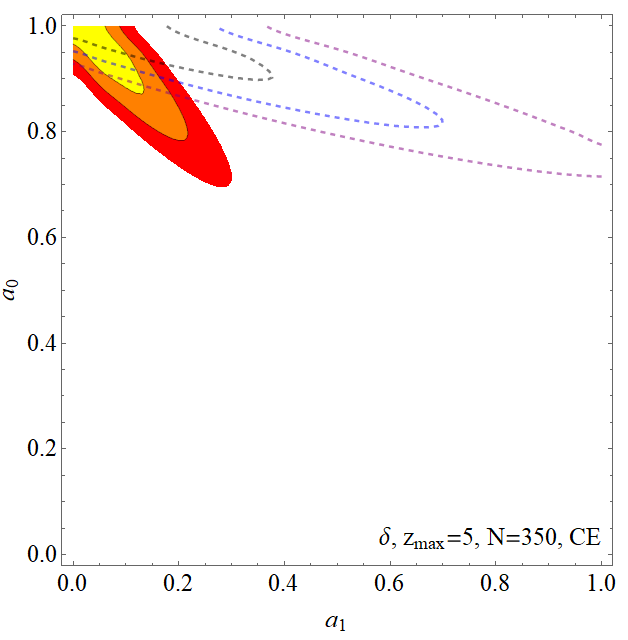}
	\end{subfigure}
	\hfill
	\begin{subfigure}{0.48\textwidth}
		\centering
		\includegraphics[width=\textwidth]{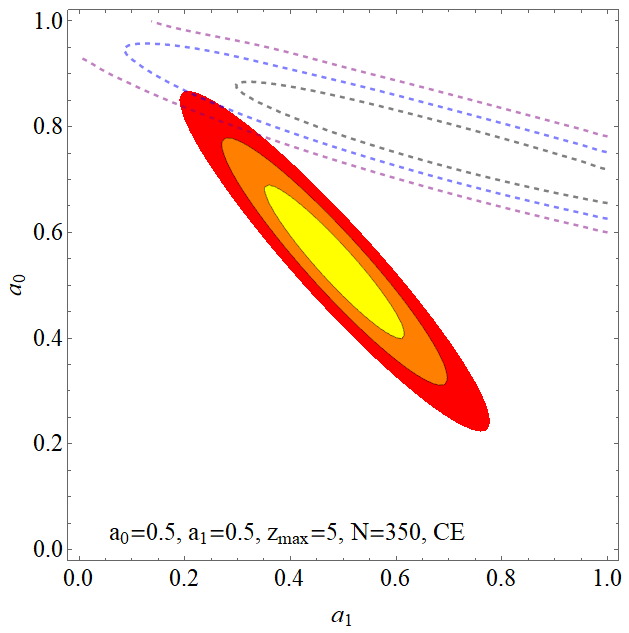}
	\end{subfigure}
	\caption{Similar constraints as above, but for Cosmic Explorer, for comparison.}
	\label{fig:CE_constraints}
\end{figure*}

To better quantify the differences, the mean values for our $350$ mock sources are $\rho_{ET} \simeq 100$ and $\rho_{CE} \simeq 210$. Although CE will perform slightly better, we find that both detectors will be able to put strong constraints on the inhomogeneity  parameters (Figures \ref{fig:delta_constraints} and \ref{fig:CE_constraints}) and will be able to break the degeneracy with modified gravity models, reaching an accuracy of the order of $\mathcal{O}(0.01)$ for the friction parameter $\nu$ (Figures \ref{fig:ET_CE_nu_likelihood}).
	
Therefore, we conclude that with their higher SNR, these detectors would be invaluable for cosmological studies.

\section{Conclusions}\label{sec:conclusions}

In this work, we propose the use of GWs standard sirens as a quite clean and model-independent probe for studying inhomogeneities in the universe. Modelling the inhomogeneities with two effective distance formulae, the DR and mDR, we investigate the sensitivity of future constraints on the inhomogeneity parameters. Furthermore, we investigate the degree of degeneracy between the impact of inhomogeneities and of modified gravity theories on the luminosity distance of GWs. We claim that a possible confusion on a test of gravity occurs due to the indistinguishabilty of gravity effects from inhomogeneity.

More specifically, we show that:

\begin{itemize}
	\item Constraints to effective inhomogeneity parameters are possible from future standard sirens observations. These are more dependent on the horizon redshift of the observations, than on the number of sources observed. We see that low redshift observations lead to very weak constraints on inhomogeneous models. Hence, high redshift observations ($z \geq 1.5$), are needed to provide a clearer probe  for inhomogeneities.
	\item Realistic inhomogeneities, based on numerical simulations of cosmological structure formation, lead to constraints consistent with an FLRW geometry.
	\item A modified propagation due to an inhomogeneous background can lead to constraints on the geometry of the universe itself. We have neglected any angular dependence, but since we are considering narrow beams, a possible presence of anisotropies could be directly constrained by future observations.
	\item An inhomogeneous background can ``mimic'' modified gravity models in the amplitude decay of a GW. This should be taken into account, when trying to constrain parameters in these models. Most extreme cases can be easily disentangled, but we have shown that modifications in the $\nu$ parameter, of the order of $\mathcal{O}(0.1)$ or high SNRs, would be needed to disentangle these effects from inhomogeneities. For these, future detectors \citep{LisaCounterpart, Abbott_2017_next_gen, Einstein_Telescope} are necessary. At the same time, a significant number of standard sirens ($N \geq 100 $) is necessary to avoid misidentifications. Similar care in the interpretation should be taken, when constraining other physically equivalent effects that lead to larger observed distances, like the opacity of the intergalactic medium \citep{Wei_2019, Zhou_et_al_2019}.  
	\item Future, ground-based GW detectors will be crucial for cosmological studies, putting strong constraints on the inhomogeneity parameters and breaking the degeneracy between modified gravity effects and matter anisotropies by measuring $\nu$ at $5 \%$ and $1 \%$ level with $100$ and $350$ events respectively.
\end{itemize}

\section*{Acknowledgements}

We want to thank Tjonnie Li for useful clarifications and the referee for their comments. SA thanks for the hospitality at Royal Observatory in Edinburgh during the Nagoya University and The University of Edinburgh Joint-Degree Program. SA also thanks Chul-Moon Yoo, M. Takada, and M. Oguri for fruitful comments. For the analysis we used \verb#Mathematica# \citep{Mathematica}, \verb#Matplotlib# \citep{Hunter:2007} and \verb#NumPy# \citep{oliphant2006guide, van2011numpy}. 

\section*{Data Availability}

The data underlying this article will be shared on reasonable request to the corresponding author.

\newpage

%%%%%%%%%%%%%%%%%%%%%%%%%%%%%%%%%%%%%%%%%%%%%%%%%%

%%%%%%%%%%%%%%%%%%%% REFERENCES %%%%%%%%%%%%%%%%%%

% The best way to enter references is to use BibTeX:

\bibliographystyle{mnras}
\bibliography{biblio} % if your bibtex file is called example.bib

% Alternatively you could enter them by hand, like this:
% This method is tedious and prone to error if you have lots of references
%\begin{thebibliography}{99}
%\bibitem[\protect\citeauthoryear{Author}{2012}]{Author2012}
%Author A.~N., 2013, Journal of Improbable Astronomy, 1, 1
%\bibitem[\protect\citeauthoryear{Others}{2013}]{Others2013}
%Others S., 2012, Journal of Interesting Stuff, 17, 198
%\end{thebibliography}

%%%%%%%%%%%%%%%%%%%%%%%%%%%%%%%%%%%%%%%%%%%%%%%%%%

%%%%%%%%%%%%%%%%% APPENDICES %%%%%%%%%%%%%%%%%%%%%

\appendix

\section{The effect of an inhomogeneous universe to merger rates}\label{sec:ApA}

Using a modified ``Hubble expansion'', given by:

\begin{equation}
\tilde{H}(z) = H_0 [a(z)\Omega_{m,0} (1+z)^3 + \Omega_{\Lambda, 0}+\Omega_{k,0}(1+z)^2]^{1/2},
\end{equation}

and the modified comoving distance, 

\begin{equation}
\tilde{D}_c = c \int_0^z \frac{1}{\tilde{H}(z')} dz',
\end{equation}{}

the merger rates are:

\begin{equation}
P(z) \sim \frac{4 \pi \tilde{D}^2_c(z) R(z)}{\tilde{H}(z)(1+z)}, 
\end{equation}{}

\begin{figure}
	\includegraphics[width=\columnwidth]{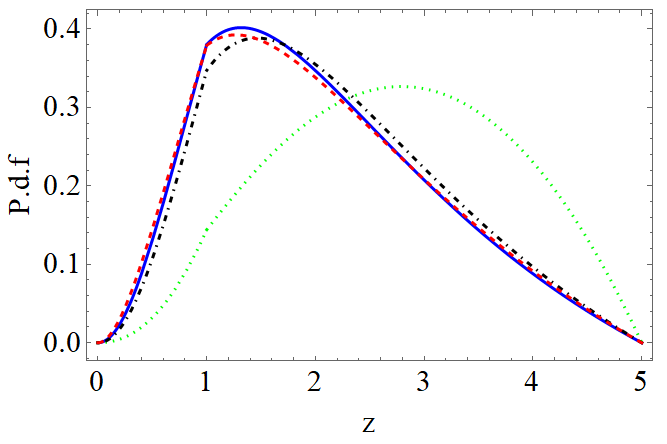}
	\caption{Dependence of probability of mergers per redshift for different cosmological parameters. The (solid, blue) line shows the FLRW result, the (green, dotted) line shows the extreme case of an ``empty beam'' ($\alpha = 0$), the (black, dot-dashed) line shows the case with  $\alpha = 0.5$ and the (red, dashed) line shows the case where ($\Omega_{\Lambda},\ \Omega_{k},\ \alpha) = (0.0,\ 0.7,\ 0.1$).}
	\label{fig:merger_rates}
\end{figure}

Although this may lead to some differences (Figure \ref{fig:merger_rates}), the effect is small. Also, we want to emphasize a caveat of this analysis: the uncertainties of the stellar population and evolution models (summarised in $R(z)$) should be more important than the effects of inhomogeneities, so the distribution of merger rates per redshift isn't a clear probe (except maybe for some very extreme cases that are not very plausible). Hence this reinforces our arguments that it's safe to assume a homogeneous background when estimating the mergers' distribution.

\section{Comparison of parameterisations}\label{sec:ApC}

In the main text we study two parameterisations proposed by \citep{Linder_1988, Bolejko_2011}:

\begin{enumerate}
	\item $\alpha(z) = a_0+a_1 z$.
	\item $\alpha(z) = 1 + D(z) \langle \delta \rangle_{1D}$,
\end{enumerate}{}

where we chose the function $D(z)$ as $D(z) = (1+z)^{-5/4}$ in order to be consistent with the weak lensing approximation \citep{Bonvin_et_al_2006}. The $\langle \delta \rangle_{1D}$ denotes the average present-time density contrast along a ray. 

Although the first one is more general, the two parameterisations are connected at small redshifts. A Taylor expansion of (ii) gives:
\begin{equation}
\alpha (z) = 1 + (1+z)^{-5/4} \langle \delta \rangle_{1D} \simeq 1 + \langle \delta \rangle_{1D} - \frac{5}{4} z \langle \delta \rangle_{1D}.
\end{equation}
With the following identifications: $a_0 = 1 + \langle \delta \rangle_{1D}$ and $a_1 = -5\langle \delta \rangle_{1D}/4$, we see how the $\alpha$ parameters are connected to inhomogeneities along the line of sight.

\section{Density distribution and ray-tracing}\label{sec:ApRays}

In the main text we calculated the distribution of mean densities along \emph{straight} trajectories (see Figure \ref{fig:Density_contrast}). Although this approximation is valid, when anisotropies are small, as a check we performed the same exercise using a ray-tracing code. The latter propagates the rays along the potential anisotropies in our cells and calculates the density values along the ``real'' trajectories\footnote{A detailed examination of our ray-tracing method will be described in a future work. Here we use it as a sanity check.}. 

\begin{figure}
	\includegraphics[width=\columnwidth]{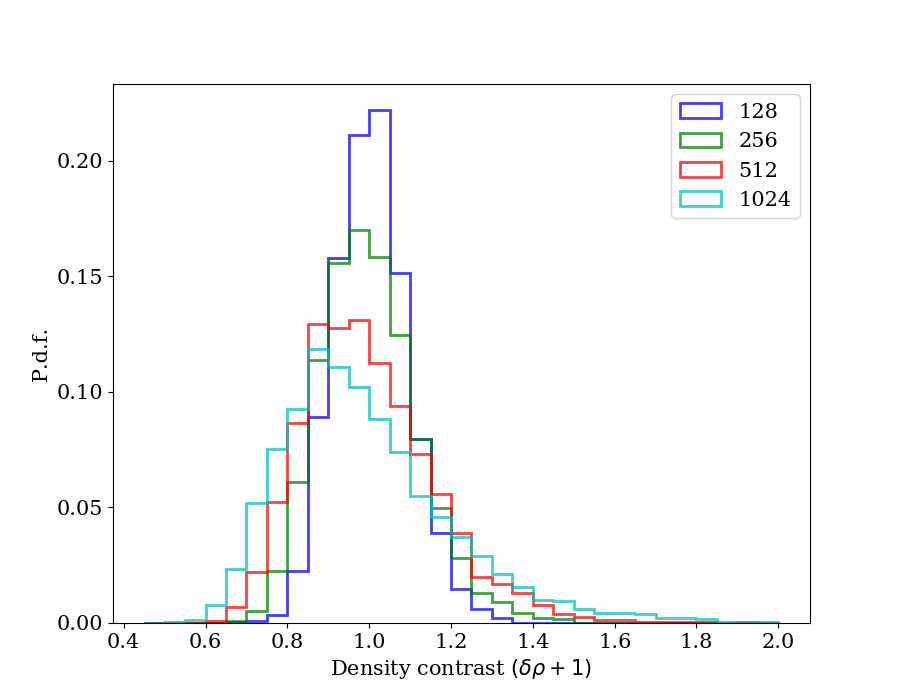}
	\caption{Distribution of densities along 1D rays $(\langle \delta \rangle_{1D} + 1)$ on simulations of different resolution, at $z=0$. The higher resolution run, which corresponds to a density averaging in $\sim 8\ \rm{Mpc}^3$ cubes gives the more interesting tails, being able to resolve better the small-scale structure.}
	\label{fig:density_rays}
\end{figure}

The results are shown in Figure \ref{fig:density_rays} and are consistent with the ones we described in the main text, validating our analysis.

\section{Degeneracy with cosmological parameters}\label{sec:ApCosmoA}

In the main text we have assumed a strong prior for the cosmological parameters, in effect fixing their values to the ones of ``concordance'' cosmology. We argued this was a valid assumption, since as previous studies \citep{Basti_et_al_2012, Fleury_et_al_2013, Dhawan_et_al_2018} have shown, the degeneracy is not strong enough to significantly change the inferred cosmological parameters, when fitting observational data for the Hubble diagram.  

In this section, we confirm these results and investigate possible future constraints from GWs observations, by fitting the inhomogeneous DR model with constant $\alpha$ and varying $\Omega_{\Lambda}$ in a \emph{flat} universe.

\begin{figure}
	\includegraphics[width=\columnwidth]{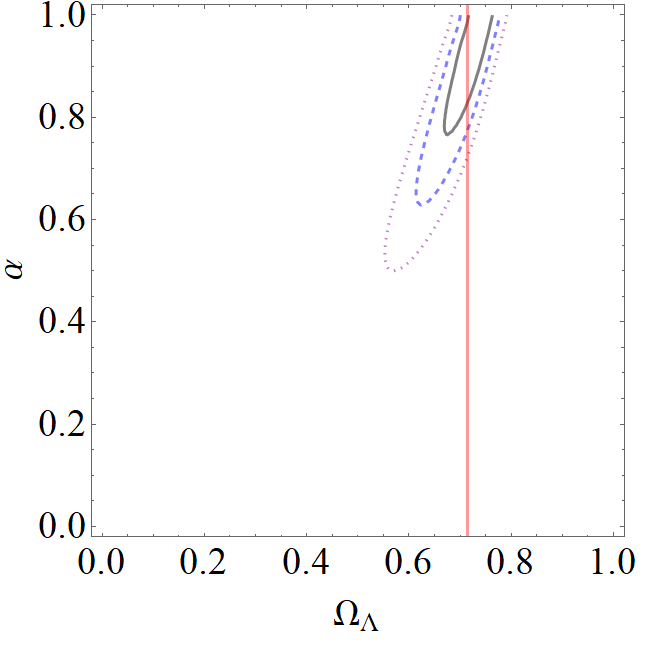}
	\caption{Constraints on ($\alpha,\ \Omega_{\Lambda}$) using a DR inhomogeneous model with constant $\alpha$, for a flat universe. We find a weak effect on $\Omega_{\Lambda}$. The red vertical line is the input value used in our simulations and the contours show ($68\%, 95.4\%, 99.7\%$) confidence intervals respectively.}
	\label{fig:cosmo_degeneracy}
\end{figure}

The results are shown in Figure \ref{fig:cosmo_degeneracy} and show a weak effect on $\Omega_{\Lambda}$, validating our analysis.

\section{Details for SNR calculation for next generation detectors}\label{sec:ApSourceSimulator}

\begin{figure}
	\includegraphics[width=\columnwidth]{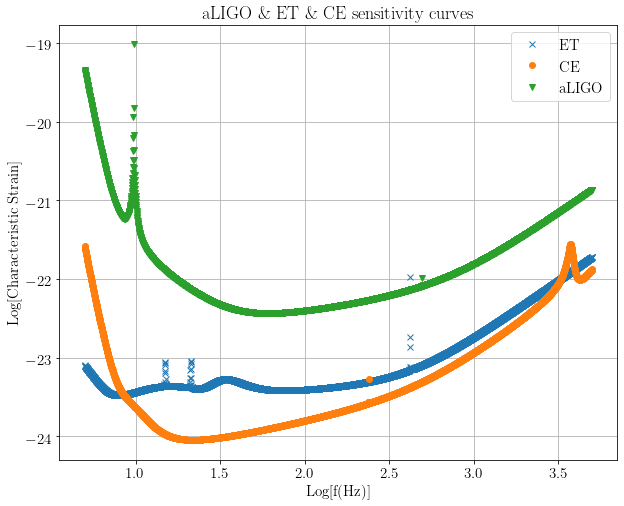}
	\caption{Sensitivity curves for aLIGO, ET and CE.}
	\label{fig:sensitivity_curves}
\end{figure}

We here show details on the SNR calculation used in this work for the next generation, ground-based detectors. Figure \ref{fig:sensitivity_curves} shows the sensitivity curves for V-shaped and L-shaped designs of ET and CE respectively \citep{Hild_et_al_2011_ETcurve, Abbott_2017_next_gen}, compared with the current aLIGO capabilities \citep{aLIGO_curve}.
	
For the generation of the primary and secondary mass we assume populations of binary neutron stars, binary black holes - neutron stars and binary black holes, with relative normalised ratios $(2/18, 1/18, 12/18)$ taken from \citep{Schneider_et_al_2001} and randomly sampled from a uniform distributions in the mass ranges  $[1, 2],\ [3,10],\ [3,50]\ M_{\sun}$ respectively, similar to \citep{Cai_et_al_2017} and \citep{LigoPopulationRates}, and impose that $M_2 \leq M_1$.

\begin{figure}
	\includegraphics[width=\columnwidth]{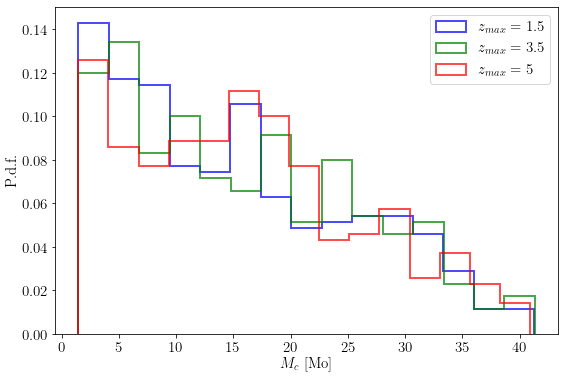}
	\caption{Distribution of chirp mass for three random samples of $350$ sources, where the maximum source redshift is $z_{\rm max} = 1.5,\ 3.5\ \& \ 5$.}
	\label{fig:Mchirp_zmax}
\end{figure}

\begin{figure}
	\includegraphics[width=\columnwidth]{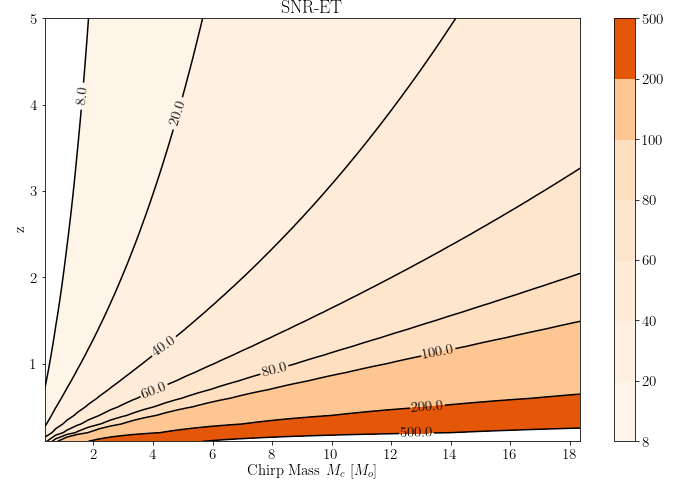}
	\caption{SNR contour plot for ET with $q=M_2/M_1=0.2$ and for $M_1$ within the range $[1, 50]\ M_{\sun}$ and the range of redshifts we have considered in the analysis above.}
	\label{fig:SNR_contour_ET}
\end{figure}

\begin{figure}
	\includegraphics[width=\columnwidth]{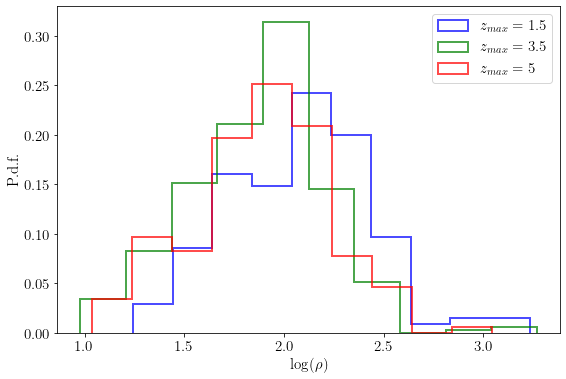}
	\caption{Distribution of the log-SNR for ET, for three random samples of $350$ sources, where the maximum source redshift is $z_{\rm max} = 1.5,\ 3.5\ \& \ 5$. The distribution moves to smaller SNRs, when higher redshifts are sampled, as expected.}
	\label{fig:SNR_zmax_ET}
\end{figure}

The distribution of chirp masses of the reference $350$ GW sources used in this work and the final SNR distribution for ET is shown in Figures \ref{fig:Mchirp_zmax} and \ref{fig:SNR_zmax_ET}. Figure \ref{fig:SNR_contour_ET} shows the SNR contour for ET, for a fixed ratio between primary and secondary mass ($q=M_2/M_1=0.2$) and for the redshifts we consider in this work. It is clear that almost all parameters lead to a signal above the detection limit.
	
\begin{figure}
	\includegraphics[width=\columnwidth]{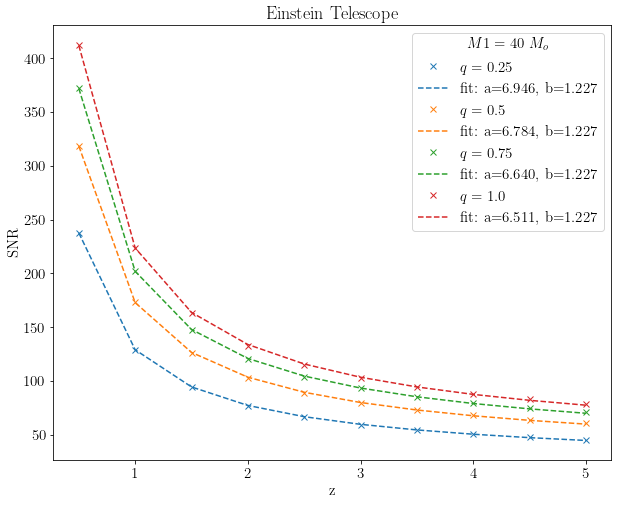}
	\caption{Dependence of SNR with redshift for ET. The primary mass is taken as $M_1 = 40\ M_{\sun}$, while the secondary takes values based on $q=M_2/M_1 = 0.25,\ 0.5,\ 0.75\ \& \ 1$. Crosses correspond to numerical points, while dashed lines are the analytical fit described in the text.}
	\label{fig:SNR_z_fit_ET}
\end{figure}

Finally, in Figure \ref{fig:SNR_z_fit_ET} we investigate the dependence of SNR with redshift for the case of ET. We numerically calculate the SNR for a range of systems and redshifts (crosses) and propose a two parameter fit:
	
\begin{equation}\label{eq:SNR_fit}
\rho = a \left(\frac{40^5}{2}\right)^{1/6} \frac{q^{1/2}}{[\frac{1}{2}(1+q)]^{1/6}} \left(\frac{M_1}{40 M_{\sun}}\right)^{5/6}\frac{(1+z)^{5/6}}{z^b}.
\end{equation}
	
\begin{figure}
	\includegraphics[width=\columnwidth]{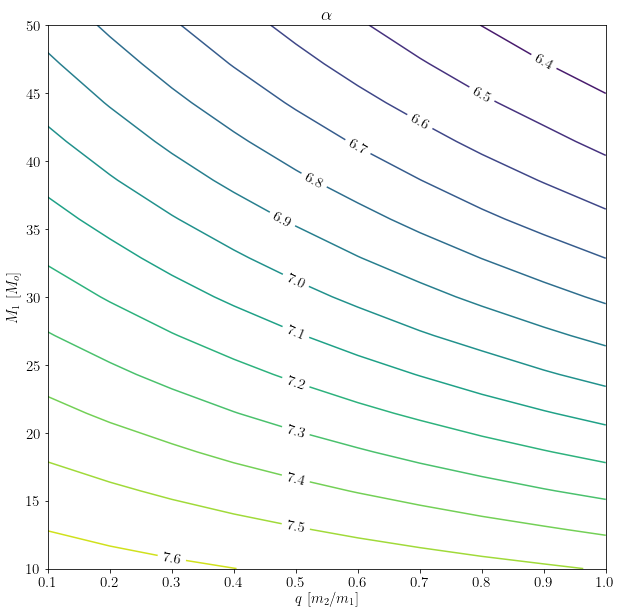}
	\caption{Dependence of the $\alpha$ parameter on the system properties. The values correspond to the ET detector.}
	\label{fig:a_values}
\end{figure}

Initial normalisation is given by $a(M_1, M_2)$ that depends on the masses of the system and the details of the sensitivity curve, while $b$ is universal and describes the redshift evolution (as parameterised this comes mostly from the luminosity distance at the denominator). With $a$ calculated numerically for a given system, i.e. for specific masses, in a specific redshift, eq. \ref{eq:SNR_fit} provides the evolution for all redshifts. Typical values for $a$ for ET are shown in Figure \ref{fig:SNR_z_fit_ET}, while a constant value of $b=1.227$ is found for all detectors and systems. The general dependence of $a$ on the masses of the system, for the ET detector, can be found in Figure \ref{fig:a_values}.
	
The value of $\rho=8$, is used as the minimum value of trigger for accepting a simulated source. This would also correspond to binary neutron stars observed, in our redshift range, with next generation, ground-based detectors, like the Einstein Telescope (ET) \citep{Einstein_Telescope} and the  Cosmic Explorer (CE) \citep{Abbott_2017_next_gen}.

\begin{table}
	\centering
	\begin{tabular}{|c|c|c|c|} 
		\hline
		& LIGO & ET & CE \\ 
		\hline \hline
		$q = 0.25$ & 0.290 & 6.946 & 14.531 \\ 
		\hline
		$q = 0.5$ & 0.272 & 6.784 & 14.278 \\
		\hline
		$q = 0.75$ & 0.254 & 6.640 & 14.015 \\
		\hline
		$q = 1$ & 0.236 & 6.511 & 13.734 \\
		\hline
	\end{tabular}
	\caption{Values of the $a$ parameter for a primary mass of $40 M_{\sun}$ and three different detectors.}
\end{table}

\begin{table}
	\centering
	\begin{tabular}{|c|c|c|} 
		\hline
		$M_1$  & $4 M_{\sun} $ & $2.1 M_{\sun}$ \\ 
		\hline \hline
		$q = 0.25$ & 7.703 & - \\ 
		\hline
		$q = 0.5$ & 7.698 & - \\
		\hline
		$q = 0.75$ & 7.691 & 7.708  \\
		\hline
		$q = 1$ & 7.682 & 7.706  \\
		\hline
	\end{tabular}
	\caption{Values of the $a$ parameter for ET and two different primary masses corresponding to smaller mass systems and binary neutron stars.}
\end{table}

\newpage

\begin{figure}
	\includegraphics[width=\columnwidth]{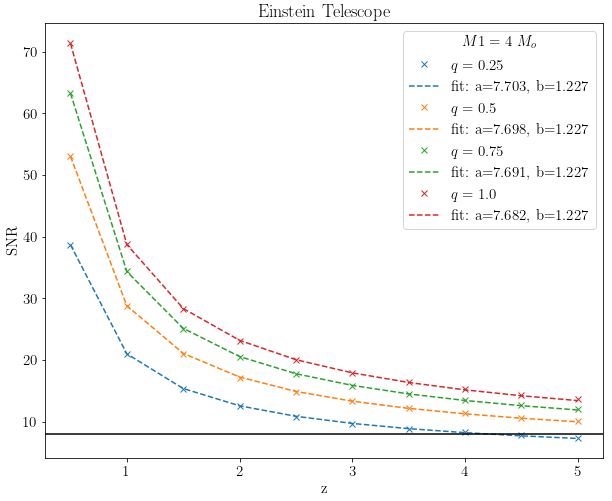}
	\caption{Dependence of SNR with redshift for ET. The primary mass is taken as $M_1 = 4\ M_{\sun}$, while the secondary takes values based on $q=M_2/M_1 = 0.25,\ 0.5,\ 0.75\ \& \ 1$. Crosses correspond to numerical points, while dashed lines are the analytical fit described in the text. The black solid line denotes the minimum trigger assumed for GW detectors $\rho=8$.}
	\label{fig:SNR_z_fit_ET_small_mass}
\end{figure}

\begin{figure}
	\includegraphics[width=\columnwidth]{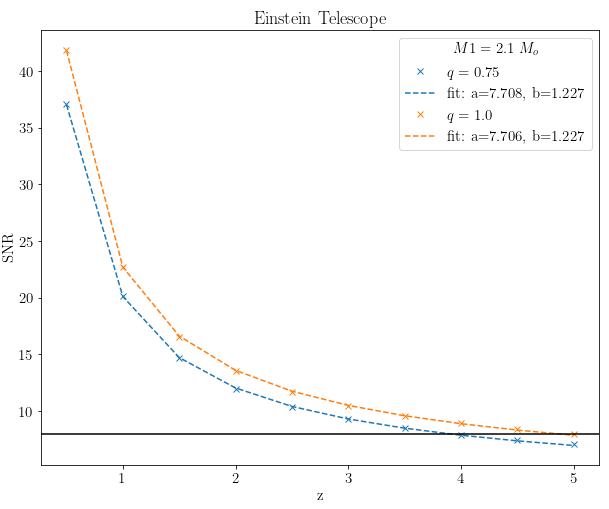}
	\caption{Dependence of SNR with redshift for ET. The primary mass is taken as $M_1 = 2.1\ M_{\sun}$, while the secondary takes values based on $q=M_2/M_1 = 0.25,\ 0.5,\ 0.75\ \& \ 1$. Crosses correspond to numerical points, while dashed lines are the analytical fit described in the text. The black solid line denotes the minimum trigger assumed for GW detectors $\rho=8$.}
	\label{fig:SNR_z_fit_ET_small_mass_NS}
\end{figure}

%%%%%%%%%%%%%%%%%%%%%%%%%%%%%%%%%%%%%%%%%%%%%%%%%%

% Don't change these lines
\bsp	% typesetting comment
\label{lastpage}
\end{document}